\documentclass[12pt]{article}

\usepackage{framed}
\usepackage{algorithm}
\usepackage[noend]{algpseudocode}
\usepackage[english]{babel}
\usepackage[utf8x]{inputenc}
\usepackage{fontenc}
\usepackage{apacite}
\usepackage{booktabs}
\usepackage{multirow}
\usepackage{siunitx}
\usepackage{mathrsfs}
\usepackage{listings}
\usepackage{color}
\usepackage{bbm}
\usepackage{adjustbox}
\usepackage{epsfig, epsf, graphicx}
\usepackage{tikz}
\usepackage{animate}
\usepackage{pstricks, pst-node, psfrag}
\usepackage{amssymb,amsmath,bm}
\usepackage{verbatim,enumerate}
\usepackage{rotating, lscape}
\usepackage{diagbox}
\usepackage[doublespacing]{setspace}
\usepackage{tikz}
\DeclareMathOperator*{\argmax}{arg\,max}
\DeclareMathOperator*{\argmin}{arg\,min}
\usetikzlibrary{arrows,calc,tikzmark}
\usepackage[colorlinks=true,linkcolor=blue,citecolor=blue]{hyperref}%
\usepackage{float}
\usepackage[small]{caption}
\usepackage{subcaption}
\usepackage[square]{natbib}
\usepackage{lipsum}
\usepackage{threeparttable}
\usepackage{multirow}
\usepackage{multicol}
\usepackage[toc,page]{appendix}

\makeatletter
\def\BState{\State\hskip-\ALG@thistlm}
\makeatother
\makeatletter
\newcounter{phase}[algorithm]
\newlength{\phaserulewidth}
\newcommand{\setphaserulewidth}{\setlength{\phaserulewidth}}

\makeatother
\setphaserulewidth{.7pt}

\newcounter{case}[algorithm]
\newlength{\caserulewidth}
\newcommand{\setcaserulewidth}{\setlength{\caserulewidth}}

\makeatother
\setcaserulewidth{.7pt}
\algdef{SE}[SUBALG]{Indent}{EndIndent}{}{\algorithmicend\ }%
\algtext*{Indent}
\algtext*{EndIndent}

\def\boxit#1{\vbox{\hrule\hbox{\vrule\kern6pt
			\vbox{\kern6pt#1\kern6pt}\kern6pt\vrule}\hrule}}

\def\bse{\begin{eqnarray*}}
	\def\ese{\end{eqnarray*}}
\def\be{\begin{eqnarray}}
\def\ee{\end{eqnarray}}
\def\bq{\begin{equation}}
\def\eq{\end{equation}}
\def\bse{\begin{eqnarray*}}
	\def\ese{\end{eqnarray*}}

\newtheorem{thm}{Theorem}
\newtheorem{defi}{Definition}
\newtheorem{assume}{Assumption}
\newtheorem{cri}{Criterion}

\begin{document}

\thispagestyle{empty} \baselineskip=28pt \vskip 5mm
\begin{center} {\Huge{\bf Robust Two-Layer Partition Clustering of Sparse Multivariate Functional Data}}
	
\end{center}

\baselineskip=12pt \vskip 10mm

\begin{center}\large
Zhuo Qu\footnote{\baselineskip=10pt Statistics Program,
King Abdullah University of Science and Technology,
Thuwal 23955-6900, Saudi Arabia.\\
E-mail: zhuo.qu@kaust.edu.sa, marc.genton@kaust.edu.sa}, Wenlin Dai\footnote[2]{\baselineskip=10pt Institute of Statistics and Big Data, Renmin University of China, Beijing 100872, China.\\
E-mail: wenlin.dai@ruc.edu.cn \\
The code and replication material is available as a supplementary material with the electronic version of this paper. } and Marc G.~Genton\textcolor{blue}{$^1$}
\end{center}
\baselineskip=17pt \vskip 10mm \centerline{\today} \vskip 15mm

\begin{center}
{\large{\bf Abstract}}
\end{center}
A novel elastic time distance for sparse multivariate functional data is proposed and used to develop a robust distance-based two-layer partition clustering method. With this proposed distance, the new approach not only can detect correct clusters for sparse multivariate functional data under outlier settings but also can detect those outliers that do not belong to any clusters. Classical distance-based clustering methods such as density-based spatial clustering of applications with noise (DBSCAN), agglomerative hierarchical clustering, and $K$-medoids are extended to the sparse multivariate functional case based on the newly-proposed distance. Numerical experiments on simulated data highlight that the performance of the proposed algorithm is superior to the performances of existing model-based and extended distance-based methods. The effectiveness of the proposed approach is demonstrated using Northwest Pacific cyclone tracks data as an example.

\baselineskip=14pt

\par\vfill\noindent
{\bf Some key words:} Cyclone data, Elastic time distance, Multivariate functional data, Outliers, Robust clustering, Sparse data

\clearpage\pagebreak\newpage \pagenumbering{arabic}
\baselineskip=26pt

\section{Introduction}
Functional data analysis (FDA, \citeauthor{ramsay2005functional} \citeyear{ramsay2005functional}) is a branch of statistics that analyzes observations that can be regarded as curves, surfaces, or any object evolving over a continuum. It has wide applications in many fields, such as meteorology, medicine, finance, biology, and language. Real-life examples of the applications of FDA include the analyses of typhoon trajectories (\citeauthor{misumi2019multivariate} \citeyear{misumi2019multivariate}), teletransmitted electrocardiograph (ECG) traces (\citeauthor{ieva2011multivariate} \citeyear{ieva2011multivariate}), stock share prices over time (\citeauthor{horvath2012inference} \citeyear{horvath2012inference}), growth curves of different body parts (\citeauthor{sheehy2000contribution} \citeyear{sheehy2000contribution}), CD4 level counts (\citeauthor{yao2005functional} \citeyear{yao2005functional}), and character handwritings (\citeauthor{kneip2000curve} \citeyear{kneip2000curve}). With the advances in data collection techniques, huge amounts of functional data are being recorded for various purposes and these data usually show heterogeneity. Hence, clustering these functional observations into homogeneous subgroups is crucial.  

Overall, functional clustering can be achieved using four types of methods, according to \citeauthor{jacques2014functional} (\citeyear{jacques2014functional}): 1) raw data methods that involve clustering directly the curves on their finite set of points---for example, the piecewise constant nonparametric density estimation by \citeauthor{boulle2012functional} (\citeyear{boulle2012functional}); 2) filtering methods that require smoothing curves into a basis of functions and clustering the resulting expansion coefficients in turn---for example, the principal points of curves (\citeauthor{tarpey2003clustering} \citeyear{tarpey2003clustering}), B-spline fitting (\citeauthor{abraham2003unsupervised} \citeyear{abraham2003unsupervised}), and functional principal component analysis (\citeauthor{peng2008distance} \citeyear{peng2008distance}); 3) adaptive methods that perform the simultaneous clustering and expression of the curves into a finite dimensional space---for example, basis expansion coefficients modelling (\citeauthor{james2003clustering} \citeyear{james2003clustering}, \citeauthor{same2011model} \citeyear{same2011model}, \citeauthor{giacofci2013wavelet} \citeyear{giacofci2013wavelet}), and functional principal component analysis (FPCA, \citeauthor{chiou2007functional} \citeyear{chiou2007functional}, \citeauthor{bouveyron2011model} \citeyear{bouveyron2011model}, \citeauthor{jacques2013funclust} \citeyear{jacques2013funclust}, \citeauthor{jacques2014model} \citeyear{jacques2014model}, \citeauthor{centofanti2021sparse} \citeyear{centofanti2021sparse}); and 4) distance-based methods where usual clustering algorithms are applied with specific distances for functional data. For example, the $K$-means algorithm (\citeauthor{hartigan1979algorithm} \citeyear{hartigan1979algorithm}) was recommended by \citeauthor{cuesta2007impartial} (\citeyear{cuesta2007impartial}), \citeauthor{antoniadis2013clustering} (\citeyear{antoniadis2013clustering}), \citeauthor{garcia2015k} (\citeyear{garcia2015k}) and \citeauthor{albert2021band} (\citeyear{albert2021band}); an agglomerative hierarchical clustering algorithm (\citeauthor{day1984efficient} \citeyear{day1984efficient}) was considered by \citeauthor{liu2012correlation} (\citeyear{liu2012correlation}); and the $K$-medoids method (\citeauthor{kaufman1990partitioning} \citeyear{kaufman1990partitioning}), which is a modification of the $K$-means method, was developed by \citeauthor{chen2017delineating} (\citeyear{chen2017delineating}); in addition, \citeauthor{floriello2017sparse} (\citeyear{floriello2017sparse}) extended the $K$-means, $K$-medoids, and hierarchical clustering algorithms to functional frameworks. Although the FPCA-based clustering method in \citeauthor{centofanti2021sparse} (\citeyear{centofanti2021sparse}) can be applied to an irregular time grid, there may be computation issues due to the estimation of parameter sets in big functional data clustering. 

For multivariate functional data, however, only few innovations exist for data clustering. From the viewpoint of generalization of adaptive methods, \citeauthor{jacques2014model} (\citeyear{jacques2014model}) generalized functional clustering from univariate to multivariate functional data via multivariate functional principal component analysis (MFPCA, \citeauthor{ramsay2005functional} \citeyear{ramsay2005functional}). Furthermore, \citeauthor{schmutz2020clustering} (\citeyear{schmutz2020clustering}) extended the work of \citeauthor{jacques2014model} (\citeyear{jacques2014model}), and their new method (funHDDC) is advantageous from two perspectives: for modelling all principal component scores whose estimated variances are non-null and for using the expectation-maximization (EM) algorithm to propose a criterion for selecting the number of clusters. Moreover, \citeauthor{misumi2019multivariate} (\citeyear{misumi2019multivariate}) proposed a multivariate nonlinear mixed-effects model to express multivariate functional data and applied a nonhierarchical clustering algorithm based on self-organizing maps to the predicted coefficient vectors of individual-specific random effect functions. With regard to distance methods, \citeauthor{ieva2011multivariate} (\citeyear{ieva2011multivariate}), \citeauthor{ieva2013multivariate} (\citeyear{ieva2013multivariate}), and \citeauthor{meng2018new} (\citeyear{meng2018new}) proposed new distance measures and applied the $K$-means clustering algorithm to multivariate functional data. Other methods considered clustering curves while capturing the features of amplitude or phase variations (\citeauthor{10.1214/14-EJS901} \citeyear{10.1214/14-EJS901}). For instance, \citeauthor{sangalli2010k} (\citeyear{sangalli2010k}) proposed a $K$-means clustering method to decompose amplitude and phase variations while assuming a linear warping time function; \citeauthor{park2017clustering} (\citeyear{park2017clustering}) proposed a conditional subject-specific warping framework and considered multivariate functional clustering with phase variation. 
  
There are criteria to find the optimal number of clusters, i.e., $K$, for distance-based clustering methods, such as $K$-means, $K$-medoids, agglomerative hierarchical clustering, and density-based spatial clustering of applications with noise (DBSCAN, \citeauthor{ester1996density} \citeyear{ester1996density}). Generally, the optimal $K$ can be determined from the elbow value (\citeauthor{shi2021quantitative} \citeyear{shi2021quantitative}), average silhouette (\citeauthor{struyf1997clustering} \citeyear{struyf1997clustering}, \citeauthor{batool2021clustering} \citeyear{batool2021clustering}), gap statistics (\citeauthor{tibshirani2001estimating} \citeyear{tibshirani2001estimating}), model-based Akaike Information Criterion (AIC, \citeauthor{akaike1974new} \citeyear{akaike1974new}), Bayesian Information Criterion (BIC, \citeauthor{schwarz1978estimating} \citeyear{schwarz1978estimating}), and integrated classification likelihood (\citeauthor{biernacki2000assessing} \citeyear{biernacki2000assessing}). 

Two main characteristics are observed for the existing clustering methods: first, for sparse multivariate functional data, no specific distance measurements are defined and the model-based clustering methods, except for that of \citeauthor{misumi2019multivariate} (\citeyear{misumi2019multivariate}), cannot be applied directly; second, the above methods rarely consider noises or outliers, though real data are often corrupted by them, see \citeauthor{ronchetti2021main} (\citeyear{ronchetti2021main}) for a recent review. Robust clustering methods for multivariate data to reduce the influence of outliers are acknowledged, see density-based spatial clustering of applications with noise (DBSCAN, \citeauthor{ester1996density} \citeyear{ester1996density}), robust spectral clustering (\citeauthor{li2007noise} \citeyear{li2007noise}), improvements in $K$-means clustering (\citeauthor{wang2011improved} \citeyear{wang2011improved}) and $K$-medoids clustering (\citeauthor{al2014novel} \citeyear{al2014novel}), and hierarchical clustering (\citeauthor{balcan2014robust} \citeyear{balcan2014robust}, \citeauthor{gagolewski2016genie} \citeyear{gagolewski2016genie}). In addition, developments in outlier detection methods in multivariate functional data are, for instance, functional adjusted outlyingness and central-stability plots (\citeauthor{hubert2015multivariate} \citeyear{hubert2015multivariate}), magnitude-shape plot (\citeauthor{dai2018multivariate} \citeyear{dai2018multivariate}), and directional outlyingness (\citeauthor{dai2019directional} \citeyear{dai2019directional}). However, robust clustering algorithms that correctly obtain clusters and detect outliers for sparse multivariate functional data are lacking. 

Sparse (multivariate) functional data are defined as data objects with various time grids per subject. One common example of sparse data in practice is imbalanced data, where some objects may have a large number of measurements while others may have few measurements. Here, the measurement is assumed to be either observed or missing for all variables. The case where the measurements for one index are observed for some variables but missing for others is not considered. In addition, the location of missing observations is assumed to be random, and independent of the observed and missing values.

The objective of this work is to develop a method that can correctly detect clusters and flag outliers, if any, for sparse multivariate functional data. First, the concept of elastic time distance (ETD) is proposed, which is applicable to (multivariate) functional data with either identical or different time measurements per subject. Second, an outlier-resistant method to obtain clusters and remove outliers based on the ETD measures is presented. The rest of the paper is organized as follows. Section \ref{sec2} proposes a novel robust clustering method based on a new distance measure, ETD. The proposed method is referred to as robust two-layer partition (RTLP) clustering. Section \ref{sec3} describes numerical experiments performed to evaluate the clustering precision and outlier detection performance among RTLP clustering, one existing model-based method, and three distanced-based methods extended with ETD. Section \ref{sec4} presents the results of application of RTLP clustering to Northwest Pacific cyclone tracks data to establish its efficiency. Section \ref{sec5} concludes the paper with a summary and discussion. Proofs are collected in the Appendix.

\section{Multivariate Functional Data Clustering}
\label{sec2}
A multivariate functional random variable of $p$ dimensions
is a $p$-variate random vector with values in an infinite-dimensional space. In a well-known model of multivariate functional data (\citeauthor{hsing2015theoretical} \citeyear{hsing2015theoretical}), the data are considered as sample paths of a stochastic process $\bm{Y}=\{\bm{Y}(t):=(Y^{(1)}(t),\ldots, Y^{(p)}(t))^\top\}_{t \in \mathcal{T}}$ taking real values in some Hilbert space, $\mathcal{H}$, of functions defined on some compact set $\mathcal{T} \subseteq \mathbb{R}$. Let $\mathcal{H}$ be square integrable functions of $\mathcal{T}$, written as $\mathcal{H}:= L^2(\mathcal{T})\times \cdots \times L^2(\mathcal{T})$. 
 
\subsection{Elastic Time Distance}
The distance measure between $\bm{Y}_m$ and $\bm{Y}_n$ for $\bm{Y}_m, \bm{Y}_n \in \mathcal{C}^p(\mathcal{T})$ is introduced, where $\mathcal{C}^p(\mathcal{T})$ represents the $p$-vectors of continuous functions on $\mathcal{T}$. Suppose Assumption \ref{time} below holds.
\begin{assume} Let the time/wavelength points come from a design density $g(u)$ such that $G(t)=\int_{-\infty}^{t}g(u){\rm d}u$. Let the compact set $\mathcal{T}=[G^{-1}(0), G^{-1}(1)]$. It is assumed that $g$ is differentiable and $\inf\limits_{t \in \mathcal{T}} g(t)>0$.
\label{time}
\end{assume}
Instead of the existing $L^p$ metric in the Hilbert space, $\mathcal{H}$, the supremum norm of point-wise $L^2$ norm between $\bm{Y}_m$ and $\bm{Y}_n$ is adopted to measure the similarity between $\bm{Y}_m$ and $\bm{Y}_n$:
\begin{equation}
d(\bm{Y}_m,\bm{Y}_n) = \sup\limits_{t \in \mathcal{T}} \Biggl[\sqrt{\sum_{l=1}^p \Big\{Y_m^{(l)}(t)-Y_n^{(l)}(t)\Big\}^2}\Biggl].
\label{d}
\end{equation}
Overall, $d(\bm{Y}_m, \bm{Y}_n)$ is an extended Chebyshev distance in the Hilbert space, with the point-wise distance being the $L^2$ norm for $p$-dimensional vectors. Other norms, such as $L^1$ norm and $L^\infty$ norm, can be adopted as a point-wise distance measure in $d(\bm{Y}_m, \bm{Y}_n)$ in principle. With the $L^2$ norm as an example, $d(\bm{Y}_m,\bm{Y}_n)$ is proved to be a metric (\citeauthor{Kelley1955general} \citeyear{Kelley1955general}, p.~119) satisfying the properties described in Theorem \ref{distan_thm}.
\begin{thm} For $\bm{Y}_m, \bm{Y}_n, \bm{Y}_o \in \mathcal{C}^p(\mathcal{T})$, d, as a metric, is a distance function:

1. Non-negativity: $d(\bm{Y}_m,\bm{Y}_n) \geq 0$.

2. Non-degeneracy: $d(\bm{Y}_m,\bm{Y}_n)=0 \iff \bm{Y}_m=\bm{Y}_n$.

3. Symmetry: $d(\bm{Y}_m,\bm{Y}_n)=d(\bm{Y}_n,\bm{Y}_m)$. 

4. Triangular inequality: $d(\bm{Y}_m,\bm{Y}_n) \leq d(\bm{Y}_m,\bm{Y}_o)+d(\bm{Y}_o,\bm{Y}_n)$.
\label{distan_thm}
\end{thm}

Although the observations are supposed to be infinite-dimensional, in practice, only a finite number of discrete observations $\widehat{\bm{Y}}_{n,k}$ ($n=1,\ldots, N$) of each sample path $\widehat{\bm{Y}}_n$ are evaluated at the set $\bm{t}_n=\{t_{n,k}:=G^{-1}(k/T_n), k = 1,\ldots, T_n\}$. When the set of finite observations $S:=\{\widehat{\bm{Y}}_1(\bm{t}_1), \ldots, \widehat{\bm{Y}}_N(\bm{t}_N)\}$ has the cardinality $N$, $d$ usually cannot be applied to calculate the distance between different $\widehat{\bm{Y}}_m, \widehat{\bm{Y}}_n \in S$ due to various grid settings. To make $d$ applicable to objects with different counts and locations of time points, a criterion to build a standard grid and interpolate observations at the standard grid for all curves is proposed. Then, the distance between $\widehat{\bm{Y}}_m$ and $\widehat{\bm{Y}}_n$ is approximated by the distance between the interpolated curves $\widetilde{\bm{Y}}_m$ and $\widetilde{\bm{Y}}_n$. The procedures are explained as follows and the derived distance is coined the elastic time distance (ETD) for sparse multivariate functional data.
\begin{defi} (Standard grid $\bm{st}$).~The standard grid $\bm{st}$ is defined as a set $\{st_k:=\frac{k-1}{T-1},~k = 1, \ldots, T\}$ with $T= \max\limits_{n=1}^N T_n$. 
\label{standard_grid}
\end{defi}

\begin{defi} (Interpolated observation at $\bm{st}$). The interpolation $\widetilde{\bm{Y}}_{n}$ is defined as a set $\{\widehat{\bm{Y}}_{n}(\widetilde{t}_{n,k}), \widetilde{t}_{n,k}: =\argmin_{t \in \bm{t}_n}|t-{st}_k|,~k = 1,\ldots, T\}$. 
\label{interpolation}
\end{defi}
\begin{figure}[b!]
   \centering
   \includegraphics[height=7cm, width=0.8\textwidth]{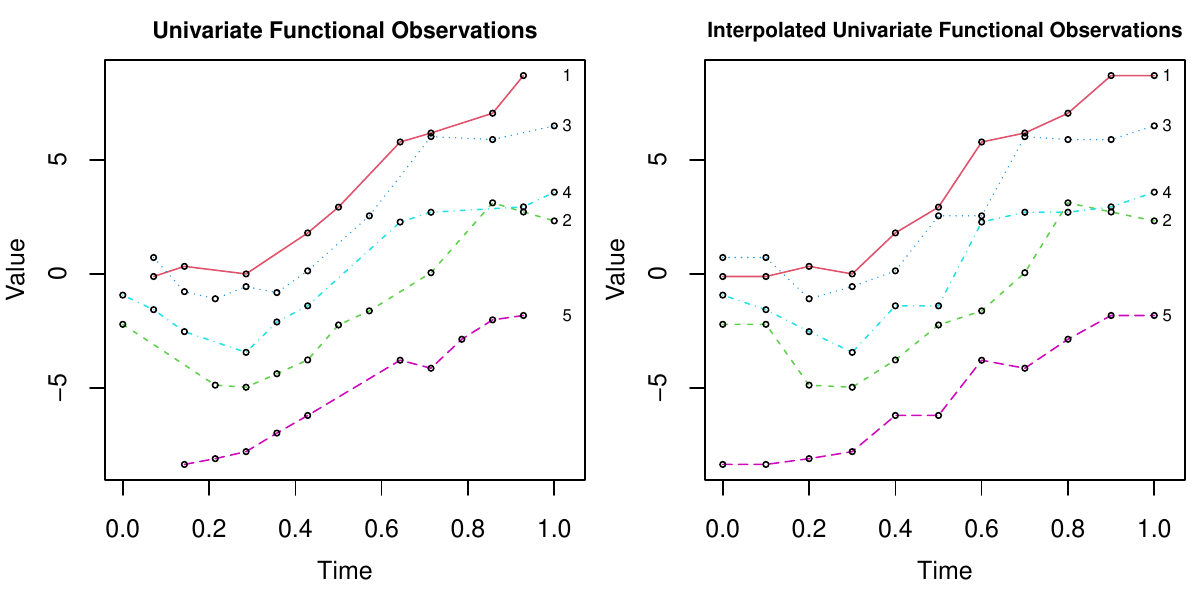}
\caption{Example of five univariate functional samples evaluated on various time grids (left), and the corresponding interpolated univariate functional samples evaluated on the standard grid (right).}
\label{etd_figure}
\end{figure}

As shown in Figure \ref{etd_figure}, because the maximum number of measurements $T$ from the five samples $\{\widehat{\bm{Y}}_1,\ldots, \widehat{\bm{Y}}_5\}$ is 11, $\bm{st}=\{(k-1)/10, k=1, \ldots, 11\}$. Then, the observation at any standard time $st_k~(k = 1, \ldots, 11)$ is the one with the closest time difference from $\bm{t}_n$ and $st_k$. Although the trend of the interpolated curve is slightly different from the original one (see the different measurements in [0.4, 0.6] before and after the interpolation), Definitions~\ref{standard_grid}-\ref{interpolation} provide a simple and feasible criterion in interpolations at the standard time grid. The interpolated curve is used to approximate the original curve and capture the similarity of the two curves with the farthest point-wise distance over the standard grid. Hence this simple interpolation criterion makes it feasible and effective to measure the similarity between curves.

For any $\widehat{\bm{Y}}_m$, $\widehat{\bm{Y}}_n \in S$ ($1 \leq n \leq m \leq N$), the finite-grid ETD between $\widehat{\bm{Y}}_m$ and $\widehat{\bm{Y}}_n$ is calculated as follows:
\begin{equation}
\begin{split}
ETD(\widehat{\bm{Y}}_m,\widehat{\bm{Y}}_n) := d(\widetilde{\bm{Y}}_m, \widetilde{\bm{Y}}_n) &= \max\limits_{k=1,\ldots,T-1}\Biggl(\max\limits_{t \in [st_k, st_{k+1}]}\Biggl[\sqrt{\sum_{l=1}^p\Big\{\widetilde{Y}^{(l)}_{m}(t)-\widetilde{Y}^{(l)}_{n}(t)\Big\}^2}\Biggr]\Biggr)\\
& =\max\limits_{k=1,\ldots,T}\Biggl[\sqrt{\sum_{l=1}^p\Big\{\widetilde{Y}^{(l)}_{m}(st_k)-\widetilde{Y}^{(l)}_{n}(st_k)\Big\}^2}\Biggr].
\label{dist}
\end{split}
\end{equation}
When all the curves are from common and equidistant time points, Definitions \ref{standard_grid}-\ref{interpolation} make no difference between $\widehat{\bm{Y}}_m$ and $\widetilde{\bm{Y}}_m~(m = 1, \ldots, N)$. Then, $d(\widetilde{\bm{Y}}_m, \widetilde{\bm{Y}}_n)=d(\widehat{\bm{Y}}_m, \widehat{\bm{Y}}_n)$. When various time grids are allowed, $ETD(\widehat{\bm{Y}}_m, \widehat{\bm{Y}}_n)$ is a pseudo-metric (\citeauthor{Kelley1955general} \citeyear{Kelley1955general}, p.~119). That is, Theorem~\ref{d} is satisfied except that $ETD(\widehat{\bm{Y}}_m, \widehat{\bm{Y}}_n)=0$ does not lead to $\widehat{\bm{Y}}_m=\widehat{\bm{Y}}_n$. If $ETD(\widehat{\bm{Y}}_m, \widehat{\bm{Y}}_n)=0$, then $\widetilde{\bm{Y}}_m=\widetilde{\bm{Y}}_n$ is obtained, but the original functions $\widehat{\bm{Y}}_m$ and $\widehat{\bm{Y}}_n$ may vary in terms of the location and number of measurement grids.

The consistency from the finite grid elastic time distance to the continuous version is satisfied given Assumption \ref{time}; see Theorem \ref{consistency} below.

\begin{thm} (Consistency).
Given $\widehat{\bm{Y}}_m, \widehat{\bm{Y}}_n \in S$, $\widetilde{\bm{Y}}_m, \widetilde{\bm{Y}}_n \in \mathcal{C}^p(\mathcal{T})$ from Definitions \ref{standard_grid}-\ref{interpolation}. Suppose Assumption \ref{time} holds. When $T_n \to \infty$ $\forall n=1, \ldots, N$, $|d(\widetilde{\bm{Y}}_m, \widetilde{\bm{Y}}_n)-d({\bm{Y}}_m, {\bm{Y}}_n)| \xrightarrow{a.s.}{} 0$.
\label{consistency}
\end{thm} 

The ETD provides a simple transformation that measures the similarity between available curves quickly. The procedure of calculating ETD in Eqn (\ref{dist}) is summarized in Algorithm~\ref{distan_alg}. The existing distance-based methods, such as the DBSCAN, agglomerative hierarchical clustering, and classical $K$-medoids, can be naturally combined with ETD. Then a distance-based clustering method is proposed with $ETD(\widehat{\bm{Y}}_m, \widehat{\bm{Y}}_n)$ as the measure in the numerical simulations. The point-wise $L^2$ norm, $L^1$ norm, and $L^\infty$ norm all achieve identically excellent performances in the simulation trials yet to be reported, and $\bm{D}$ from Algorithm \ref{distan_alg} is used to represent the result of the distance matrix.

\begin{algorithm}[h!]
\caption{Elastic Time Distance Algorithm}
\begin{algorithmic}[1]
\State \textbf{Input:} $S = \{\widehat{\bm{Y}}_1,\ldots, \widehat{\bm{Y}}_N\}$, $\bm{t}_1,\ldots, \bm{t}_N$
\State \textbf{Initialize:} Zero matrix $\bm{D} \in \mathbb{R}^{N \times N}$, $T = \max\limits_{n=1}^{N}{T_n}$, $\widetilde{\bm{Y}}_n \gets {\rm Definition}$~\ref{interpolation} $\forall~n = 1,\ldots,N$
\Function {$ETD$}{$\widehat{\bm{Y}}_i, \widehat{\bm{Y}}_j$}
\Indent
\State \textbf{return} $\max\limits_{k =1}^{T}\bigg\{\sqrt{\sum_{l=1}^{p} (\widetilde{Y}^{(l)}_{i,k}-\widetilde{Y}^{(l)}_{j,k})^2}\bigg\}$
\EndIndent
\EndFunction
\For {$i=1$ to $N-1$}
\For {$j = i+1$ to $N$}
\State $D_{i,j} \gets ETD(\widehat{\bm{Y}}_i, \widehat{\bm{Y}}_j)$
\State $D_{j,i} \gets D_{i,j}$
\EndFor
\EndFor
\State \textbf{Output:} $\bm{D}$
\end{algorithmic}
\label{distan_alg}
\end{algorithm}

\subsection{Two-Layer Partition Clustering Algorithm}
 Based on the matrix of ETD, the neighbours of each curve and the core of a set are defined, which follow the style of DBSCAN (\citeauthor{ester1996density} \citeyear{ester1996density}) 
 and depth-based clustering (\citeauthor{jeong2016data} \citeyear{jeong2016data}). First, the first-layer partition is introduced. However, the output of the first-layer partition may have disjoint sets that are from homogeneous groups. Then, another layer is implemented to make it possible to merge nearby groups into a cluster. In practice, the third-layer partition leads to one cluster that contains all curves. Hence, a two-layer partition clustering algorithm is proposed. Here, necessary concepts are introduced before describing the proposed clustering algorithm. Let $|A|$ be the cardinality of the set $A$, and $q_{\theta}(A)$ be the $\theta$-quantile $(0< \theta < 1)$ of the set~$A$, and $D(S)$ be a set containing the upper triangular elements of $\bm D$ from Algorithm~\ref{distan_alg}.

\begin{defi} (Neighbours of a curve).~The set $nbr(\bm{Y}, S, \theta):= \{\bm{Z} \in S: ETD(\bm{Y}, \bm{Z}) < q_\theta(D(S))\}$ consists of neighbours of $\bm{Y}$ in the set $S$. 
\label{neighbour}
\end{defi}Here, $\theta$ is used to determine the number of neighbours per curve. Generally, the number of neighbours per curve is positively correlated with $\theta$. The criterion for selecting the optimal $\theta$ is introduced in Section \ref{clustering_execution}.

\begin{defi} (Core and center of the set). For any set $A$, its core is the element with the most neighbours, i.e., $core(A, \theta) = \argmax_{\bm{Y} \in A}|nbr(\bm{Y}, A, \theta)|$. If $A$ is a cluster, $core(A, \theta)$ is named as the center of $A$.
\label{core}
\end{defi}

\begin{defi} (Disjoint partitions). The disjoint partition $\bm{P} = (P^1, \ldots, P^L)$ of $S$ is defined as a separation of $S$ that satisfies: $P^i \cap P^j = \emptyset~(1 \leq i <j \leq L)$, $P^1 \cup \cdots \cup P^L = S$, and $|P^1|\geq |P^2|\geq \cdots \geq |P^L|$. Here, $L$ is the number of disjoint sets (also the cardinality of $\bm{P}$), and $L \leq N$. For a disjoint partition $\bm{P}$ of $S$, let $P^{i}$ be the $i$-th set in the partition $\bm{P}$.
\label{partition}
\end{defi}

The idea of the two-layer partition clustering algorithm is similar to the bottom-up approach in agglomerative hierarchical clustering. However, only two layers of hierarchy are implemented and not all curves are merged to one cluster in the two-layer partition method. In the first layer, each curve, as an isolated cluster, is merged into disjoint groups based on the concepts of neighbours and the core. Then, in the second layer, the first-layer disjoint groups are merged into a set of clusters. The criteria for determining the first and second layers of the partitions are proposed as follows.
\begin{cri} (The first-layer partition $\bm{G}$). The first-layer partition $\bm{G}$ from the core and its neighbours is obtained. First, $S$ is copied to the set $\widetilde{S}$, and $\bm{G}$ is initialized as the empty set, and $M=1$. While $\widetilde{S}$ is not empty, implement the following three steps: define the curve with the most neighbours in $\widetilde{S}$ as a core ($c = core(\widetilde{S}, \theta)$); define the set consisting of all its neighbours in $\widetilde{S}$ as $G^M$ ($G^M = nbr(c, \widetilde{S}, \theta)$); update $\widetilde{S}$ with $ \widetilde{S}=\widetilde{S}\setminus G^M$ and increase $M$ by $1$. Let $M$ represent the cardinality of $\bm{G}$.
\label{cri1}
\end{cri}

The intuition of Criterion \ref{cri1} is as follows. Assuming the current set is a copy of $S$, the idea of disjoint groups $\bm{G}$ is: determining the core of the current set, finding the neighbours of the core in the current set as the disjoint group in $\bm{G}$, and removing such neighbours from the current set. For $G^i, G^j \in \bm{G}~(1\leq i<j \leq M)$ in Criterion \ref{cri1}, $|G^i|\geq |G^j|$ is satisfied, where $G^i$ is the prioritized group, and $G^j$ is the subordinate group. The pairwise merging principle $merge(G^i, G^j)$ is as follows: a subordinate group $G^j$ may be merged to a prioritized group $G^i$ (i.e., $G^i:=G^i \cup G^j, G^j:=\emptyset$) if the core of $G^j$ is a neighbour of a curve from $G^i$ in $S$ (i.e., $core(G^j, \theta) \in \mathop{\cup}_{\bm{Y} \in G^i}nbr(\bm{Y}, S, \theta)$), and if $G^j$ was not merged to any previous prioritized group $G^k~(1\leq k<i)$~(i.e., $G^j \neq \emptyset$ 
) yet. 

The pairwise merging principle is the core of the second-layer partition $\bm{C}$. Assuming the disjoint groups $\widetilde{\bm{G}}$ is a copy of the first-layer partition $\bm{G}$, the idea of merging disjoint groups is implemented in a decreasing order. While $\widetilde{\bm{G}}$ is not empty, implement the following three steps: find the most prioritized group $\widetilde{G}^1$ in $\widetilde{\bm{G}}$; for all its subordinate groups in $\widetilde{\bm{G}}$, check whether the subordinate group is a neighbour of any curve belonging to $\widetilde{G}^1$, if yes, merge $\widetilde{G}^1$ with the subordinate group, and remove such subordinate group in $\widetilde{\bm{G}}$; include $\widetilde{G}^1$ in $\bm{C}$, and remove $G^1$ from $\widetilde{\bm{G}}$. The first-layer (group) partition is named as $\bm{G}$, and the second-layer (cluster) partition is named as $\bm{C}$, and both satisfy Definition \ref{partition}.
\begin{algorithm}[b!]
\caption{Two-Layer Partition Clustering Algorithm}
\begin{algorithmic}[1]
\State \textbf{Input}: $S = \{\widehat{\bm{Y}}_1,\ldots, \widehat{\bm{Y}}_N\}$, $D(S)$ from Algorithm \ref{distan_alg}, $\theta$
\Function {$nbr$}{$\bm{Y}, A, \theta$}
\State \textbf{return} $\{\bm{Z} \in A ~{\rm if}~ ETD(\bm{Y}, \bm{Z}) < q_{\theta}(D(S))\}$
\EndFunction
\Function {$core$}{$A, \theta$}
\State \textbf{return} $\argmax_{\bm{Y} \in A}|nbr(\bm{Y}, A, \theta)|$
\EndFunction
\Function {$merge$}{${G}^i, {G}^j$}
\If {$|{G}^i| < |{G}^j|$}
\State {\textbf{error}~~The first parameter should be the prioritized group}
\EndIf
\If {$core({G}^j, \theta) \in \mathop{\cup}_{\bm{Y} \in {G}^i} nbr(\bm{Y}, S, \theta)$}
\State {${G}^i \gets {G}^i \cup {G}^j$}
\State {${G}^j \gets \emptyset$}
\State {\textbf{return}~${G}^i, {G}^j$}
\EndIf
\EndFunction
\State \textbf{Initialize:} $\bm{G} \gets \emptyset$, $M \gets 1$, $\widetilde{S} \gets S$
\While {$\widetilde{S} \neq \emptyset$}
\State ${G}^M \gets nbr(core(\widetilde{S}, \theta), \widetilde{S}, \theta)$
\State $\widetilde{S} \gets \widetilde{S} \setminus {G}^M$
\State $M \gets M + 1$
\EndWhile
\State \textbf{Initialize:} $\bm{C} \gets \emptyset$, $I \gets 1$, $\widetilde{\bm{G}} \gets \bm{G}$
\While {$\widetilde{\bm{G}} \neq \emptyset$}
\State $\bm{R} \gets \widetilde{\bm{G}} \setminus \widetilde{{G}}^{1}$
\For {${A}$ in $\bm{R}$}
\State {$merge(\widetilde{{G}}^1, {A})$}
\EndFor
\State ${C}^{I} \gets \widetilde{{G}}^{1}$
\State $\widetilde{\bm{G}} \gets \widetilde{\bm{G}}\setminus \widetilde{{G}}^{1}$
\State $I \gets I+1$
\EndWhile
\State \textbf{Output}: $\bm{C}$
\end{algorithmic}
\label{clustalg}
\end{algorithm}

\begin{cri} (The second-layer partition $\bm{C}$). To obtain the second-layer partition $\bm{C}$, first, the partition $\bm{G}$ is copied to $\widetilde{\bm{G}}$, and $\bm{C}$ is initialized as the empty set. While $\widetilde{\bm{G}}$ is not empty, implement the following three steps: separate partition $\widetilde{\bm{G}}$ into its most prioritized group $\widetilde{{G}}^1$ and its subordinate groups $\bm{R}:=\widetilde{\bm{G}}\setminus \widetilde{{G}}^1$; for ${A} \in \bm{R}$, implement $merge(\widetilde{{G}}^1, {A})$; include $\widetilde{{G}}^1$ in $\bm{C}$, and update $\widetilde{\bm{G}}$ as $\widetilde{\bm{G}}=\widetilde{\bm{G}}\setminus \widetilde{{G}}^1$. Let $I$ represent the cardinality of the partition $\bm{C}$.
\label{cri2}
\end{cri}

Algorithm \ref{clustalg} is used to obtain clusters from two layers of hierarchy according to Definitions~\ref{standard_grid}-\ref{partition} and Criteria \ref{cri1}-\ref{cri2}.

\subsection{Cluster and Outlier Recognition Algorithm}
 The output of Algorithm \ref{clustalg} may include clusters with a large number of curves and some with few curves because of the noise and outliers in the observations. To solve this issue, Algorithm \ref{outalg} is proposed for recognizing the primary clusters and outliers using $p_m$ and $\alpha$. Here, $p_m$ denotes the minimal required number of observations in a cluster divided by the sample size, and $\alpha$ ($0<\alpha<1$) is a threshold that defines the $\alpha$-quantile of the distances between curves inside a cluster and the cluster center. Let $\Psi_x(A)$ be the empirical cumulative distribution of $x$ in the set $A$. The criterion for selecting the primary clusters and outliers is proposed as follows.

\begin{cri} (The set of primary clusters $\bm{C}_p$ and the outlier set ${O}$). The set of primary clusters is initialized as $\bm{C}_p=\{{C} \in \bm{C}$: $|{C}| \geq Np_m\}$. Naturally, the potential outlier set is ${O_p} = \{\bm{X} \in {C}$: $|{C}|< Np_m~{\rm and}~C \in \bm{C}\}$. Let the outlier set ${O}$ be empty. The potential outlier $\bm{X} \in O_p$ is labeled as an outlier ($\bm{X} \in O$), if for any $C \in \bm{C}_p$, the ETD between $\bm{X}$ and $core(C, \theta)$ is larger than the $\alpha$-quantile of the set $D(C, core(C, \theta))$, which is the set of the ETD between each element in $C$ and $core(C, \theta)$. Otherwise, $\bm{X}$ is classified into the set with minimum distance to the center distribution, ${C}^l_p = \argmin_{C \in \bm{C}_p} \Psi_{ETD(\bm{X}, core(C, \theta))}(D(C, core(C, \theta)))$.
In a mathematical notation, $\forall \bm{X} \in O_p$,
$$
\bm{X} \in
\begin{cases}
 O, & \textit{if~} ETD(\bm{X}, c) > q_{\alpha}(D(C, c)) ~\forall~ C \in \bm{C}_p, c:=core(C, \theta), \\
 {C}^l_p, & if ~{C}^l_p:=\argmin_{C \in \bm{C}_p} \Psi_{ETD(\bm{X}, c)}(D(C, c)), c:=core(C, \theta). \\
\end{cases}
$$
\label{cri3}
\end{cri}
\vspace{-1cm}
\quad Here, each element in $\bm{C}_p$ is a set containing at least $Np_m$ number of curves, while each element in $O$ is a detected outlier. 
\begin{algorithm}[h!]
\caption{Cluster and Outlier Recognition Algorithm}
\begin{algorithmic}[1]
\State \textbf{Input:} $S = \{\widehat{\bm{Y}}_1, \ldots, \widehat{\bm{Y}}_N\}$, $\bm{D}$ from Algorithm \ref{distan_alg}, $\bm{C}$ from Algorithm \ref{clustalg}, $p_m$, $\alpha$
\State \textbf{Initialize:} $\bm{C}_p \gets \emptyset$, $O_p \gets \emptyset$, $O \gets \emptyset$, $B \gets 0$
\For {$i$ in $1,\ldots, |\bm{C}|$}
\If {$|C^i|\geq N p_m$}
\State $B \gets B+1$
\State ${C}^{B}_p \gets C^i$
\Else 
\State $O_p \gets O_p \cup C^i $
\EndIf
\EndFor
\If {$|O_p|>0$ and $|\bm{C}_p|>0$}
\For {$\bm{X}$ in $O_p$}
\If {$ETD(\bm{X}, core(C, \theta)) > q_{\alpha}(D(C, core(C, \theta)))~\forall~ C \in \bm{C}_p$}
\State $O \gets O \cup \bm{X}$
\Else 
\State ${C}^l_p \gets \argmin_{C \in \bm{C}_p} \Psi_{ETD(\bm{X}, core(C, \theta))}(D(C, core(C, \theta)))$
\State ${C}^l_p \gets {C}^l_p \cup \bm{X}$
\EndIf
\EndFor
\EndIf
\State \textbf{Output:} $\bm{C}_p$ and $O$
\end{algorithmic}
\label{outalg}
\end{algorithm}
Algorithm \ref{outalg} applies Criterion \ref{cri3} and obtains a set of primary clusters $\bm{C}_p$ and outliers $O$. Let $p_m = 0.1$. It is recommended to use $\alpha$ between $0.85$ and $0.9$ based on the empirical simulations yet to be reported. Adoption of Algorithm~\ref{outalg} makes the two-layer partition clustering (Algorithm \ref{clustalg}) robust to outliers and removes outliers. This method is referred to as RTLP clustering.

\subsection{Clustering Execution}
\label{clustering_execution}
The clustering result may differ if different $\theta$ is applied. Thus, the optimal $\theta$ needs to be selected. Under $\theta \in [0, 1]$, implementing Algorithms \ref{clustalg}-\ref{outalg} obtains a set of primary clusters $\bm{C}_{p,\theta}=({C}^1_{p, \theta},\ldots, {C}^{B_{\theta}}_{p, \theta})$ and the outlier set $O_{\theta}$. Then, the average silhouette value to assess the goodness of clustering is implemented, which is the mean of the silhouette values (\citeauthor{rousseeuw1987silhouettes} \citeyear{rousseeuw1987silhouettes}) of all data objects. The silhouette value measures the resemblance of an object to its cluster compared to other clusters and ranges from $-1$ to $1$. Here, a high value means that the object is well matched to its own cluster and poorly matched to neighbouring clusters, and a low value indicates that the clustering configuration may have an inappropriate number of clusters. The optimal $\theta$ is the one with the largest average silhouette value (\citeauthor{batool2021clustering} \citeyear{batool2021clustering}). Thereafter, $\bm{C}_p$ and $O$ under the optimal $\theta$ are adopted.

The original silhouette value is only defined when the object $\bm{Y}$ belongs to a cluster ${C}^b_{p, \theta}$~($1\leq b \leq B_{\theta}$). That is, if $\bm{Y} \in {C}^b_{p, \theta}$, then the average distance within the cluster $a(\bm{Y}, \theta) = \frac{1}{|{C}^b_{p, \theta}|-1}\sum_{\bm{Z} \in {C}^b_{p, \theta}} ETD(\bm{Y}, \bm{Z})$, the minimum distance across the clusters $b(\bm{Y}, \theta)= \min\limits_{\substack{j \neq b \\ 1 \leq j \leq B_{\theta}}} \frac{1}{|{C}^j_{p, \theta}|}\sum_{\bm{Z} \in {C}^j_{p, \theta}}ETD(\bm{Y}, \bm{Z})$, and the silhouette value of $\bm{Y}$ is $s(\bm{Y}, \theta)=\frac{b(\bm{Y}, \theta)-a(\bm{Y}, \theta)}{\max\{a(\bm{Y}, \theta),~b(\bm{Y}, \theta)\}}$. The silhouette value is also defined when the object is in the outlier set $O_{\theta}$. That is, if $\bm{Y} \in O_{\theta}$, then $s(\bm{Y}, \theta) = 0$. It helps correctly recognize clusters and detect outliers if any.

Consider the silhouette value of a curve $\bm{Y}$ when it is correctly or wrongly detected, respectively. The first case is that the curve is an outlier ($\bm{Y} \in O$). If an outlier $\bm{Y} \in O$ is correctly detected ($\bm{Y} \in O_{\theta}$), then $s(\bm{Y}, \theta)=0$. Otherwise, if an outlier $\bm{Y} \in O$ is wrongly put into a cluster ($\bm{Y} \in C^l_{p, \theta}$ and $C^l_{p, \theta}\in \bm{C}_{p, \theta}$), usually the average distance within the cluster, $a(\bm{Y}, \theta)$, is larger than the minimum distance across the clusters, $b(\bm{Y}, \theta)$, hence $s(\bm{Y}, \theta)$ is negative. The second case is that the curve belongs to a cluster, i.e., $\bm{Y} \in C^b_{p}~(1 \leq b \leq B)$. When $\bm{Y}$ is correctly labeled in its cluster $C^b_{p, \theta}$, its silhouette value, $s(\bm{Y}, \theta)$, is positive because the minimum distance across the clusters, $b(\bm{Y}, \theta)$, is larger than its average distance within the cluster, $a(\bm{Y}, \theta)$. When $\bm{Y}$ is wrongly labeled in the outlier set $O_{\theta}$, its silhouette value, $s(\bm{Y}, \theta)$, turns to zero. When $\bm{Y}$ is wrongly labeled in any other clusters $C^w_{p, \theta}~(w \neq b, 1\leq w \leq B_{\theta})$, its silhouette value, $s(\bm{Y}, \theta)$, becomes negative because the minimum distance across the clusters, $b(\bm{Y}, \theta)$, is less than the average distance within the cluster, $a(\bm{Y}, \theta)$. The optimal $\theta$ is defined as $\theta = \argmax \limits_{\theta \in [0, 1]}\bar{s}(\theta)$, with the average silhouette value $\bar{s}(\theta)= \frac{1}{N}\sum_{k=1}^N s(\bm{Y}, \theta)$. Hence, the idea of setting zero values for the set of outliers promotes the separation between clusters and outliers.

\begin{figure}[!ht]
    \begin{subfigure}{0.33\textwidth}
     \centering
     \includegraphics[width = 0.9\linewidth, height = 5.5cm]{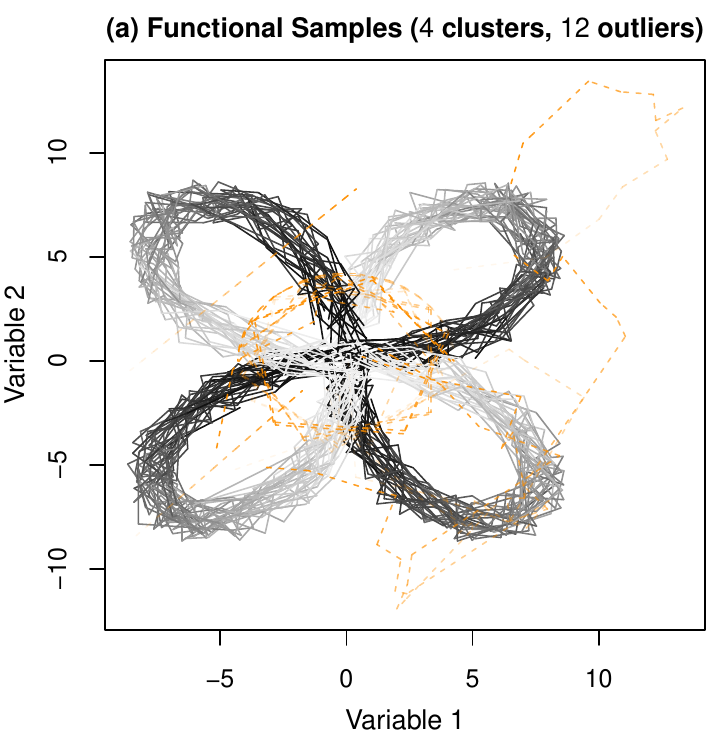}
   \end{subfigure}\hfill
    \begin{subfigure}{0.33\textwidth}
     \centering
     \includegraphics[width = 0.9\linewidth, height = 5.5cm]{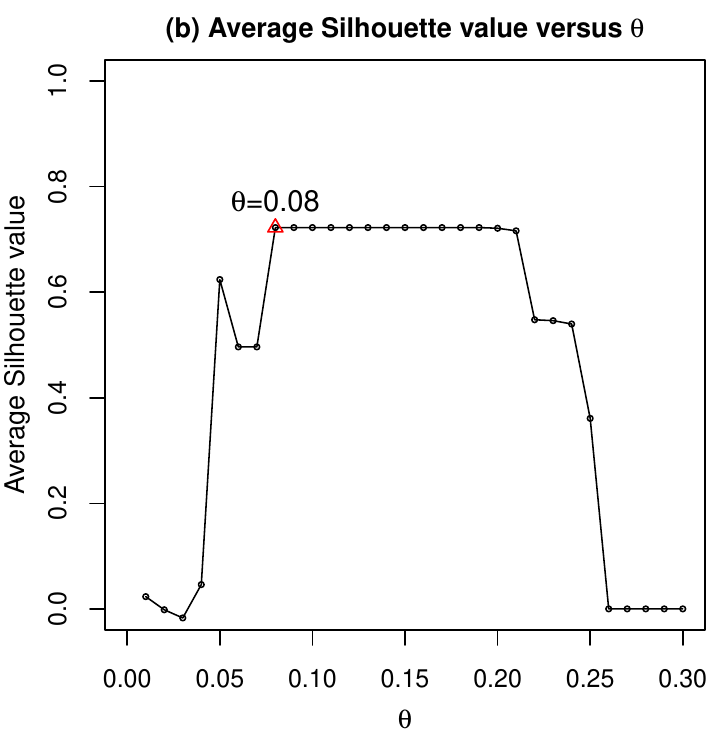}
   \end{subfigure}\hfill
   \begin{subfigure}{0.33\textwidth}
     \centering
     \includegraphics[width = 0.9\linewidth, height = 5.5cm]{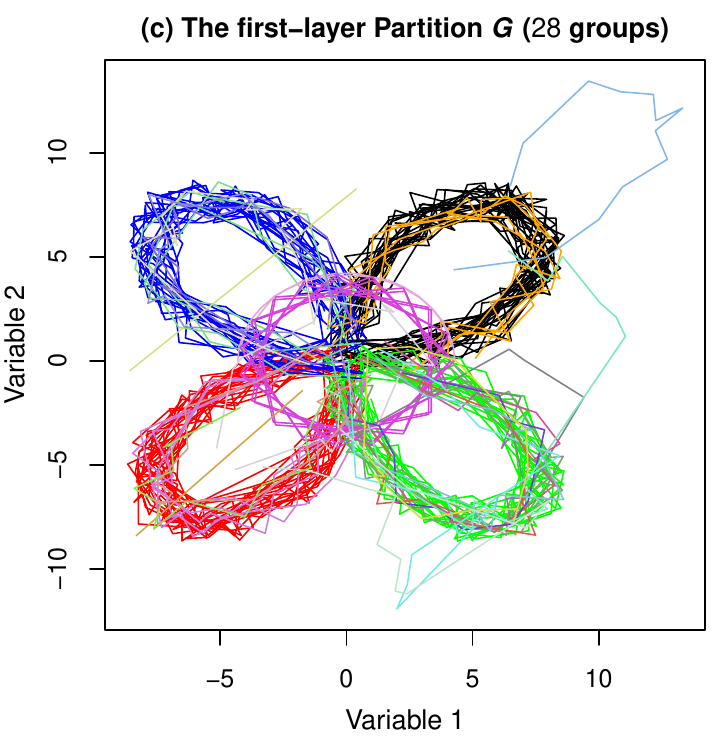}
   \end{subfigure}\hfill
   \begin{subfigure}{0.33\textwidth}
     \centering
     \includegraphics[width = 0.9\linewidth, height = 5.5cm]{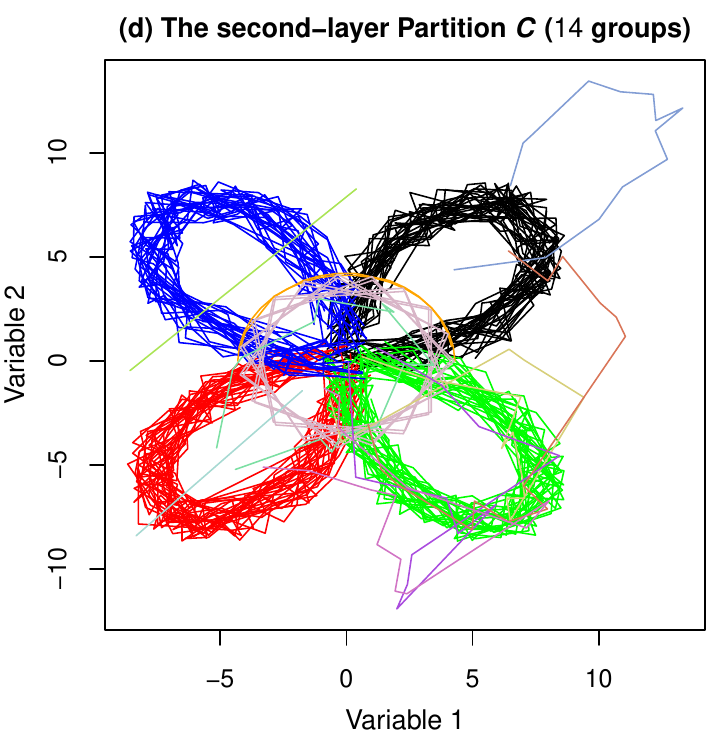}
   \end{subfigure}\hfill
   \begin{subfigure}{0.33\textwidth}
     \centering
     \includegraphics[width = 0.9\linewidth, height = 5.5cm]{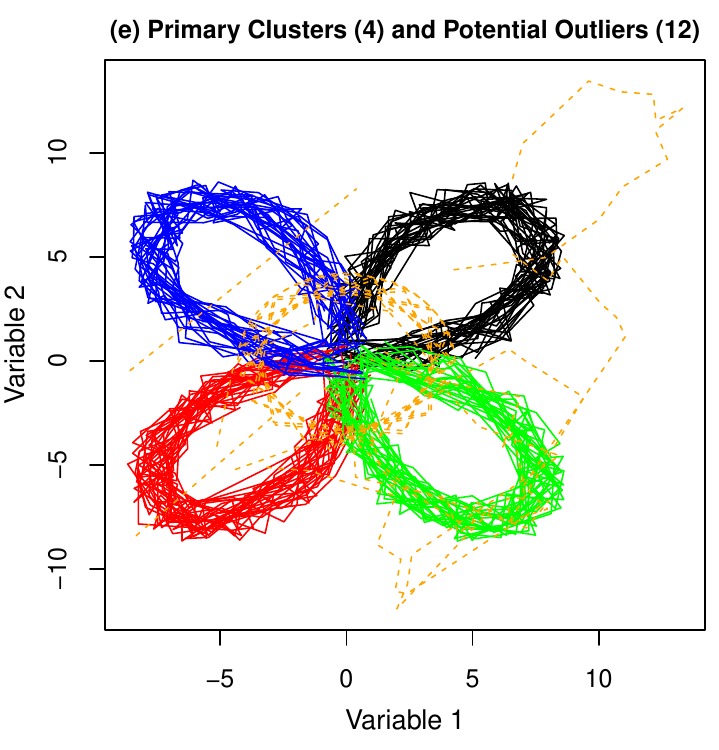}
   \end{subfigure}\hfill
   \begin{subfigure}{0.33\textwidth}
     \centering
     \includegraphics[width = 0.9\linewidth, height = 5.5cm]{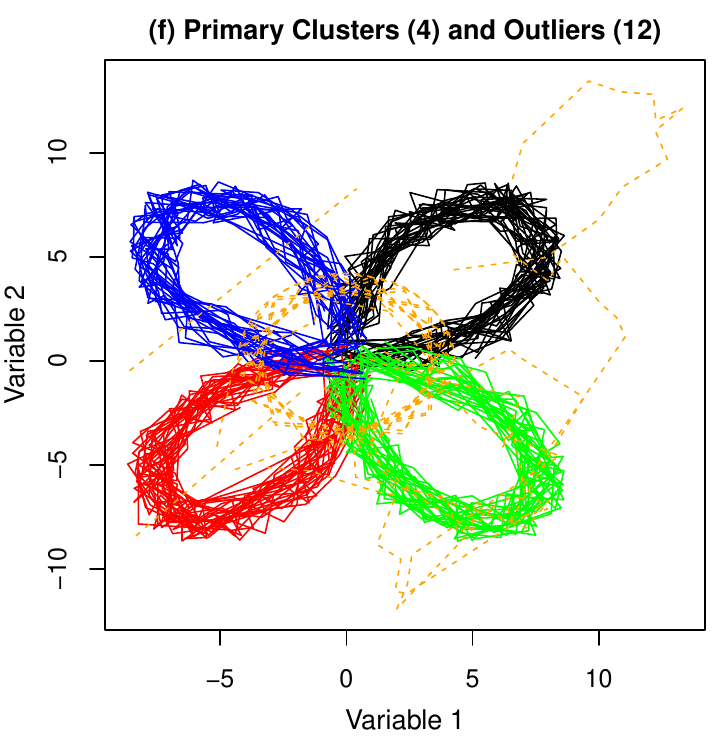}
   \end{subfigure}\hfill
    \caption{Example of stepwise clustering results. (a) Original clusters in black (transparency increases with time elapsed) mixed with outliers in orange, and (b) changes in the average silhouette value vs. $\theta$, (c) and (d) First- and second-layer partitions $\bm{G}$ and $\bm{C}$ from Algorithm \ref{clustalg}, respectively, and (e) and (f) primary clusters $\bm{C}_p$ and outliers $O$ before and after applying the detection criterion, respectively, in Algorithm \ref{outalg}.}
    \label{clustering_example}
\end{figure}

In practice, $\theta \in [0, 0.25]$, because all curves are merged into one group when $\theta \geq 0.25$ as the case in Figure \ref{clustering_example} (b). The change of average silhouette value versus $\theta$ was explored when the number of clusters changed from two to five in the simulations yet to be reported, and the average silhouette value all dropped significantly to zero when $0.2\leq \theta \leq 0.25$. 

In principle, any distance measures suitable for multivariate functional data can be used in the RTLP clustering. The ETD is used as a building block as it is applicable for both complete and sparse multivariate functional data. Overall, the RTLP clustering is executed in several steps:

1) Compute ETD according to Algorithm \ref{distan_alg}.

2) Execute Algorithms \ref{clustalg}-\ref{outalg} for $\theta \in [0.01, 0.25]$ and compute the average silhouette value $\bar{s}(\theta)$. The optimal $\theta$ is the one with the highest $\bar{s}(\theta)$.

3) Implement Algorithms \ref{clustalg}-\ref{outalg} under the optimal $\theta$ from 2), and obtain the set of primary clusters $\bm{C}_p$ and the outlier set $O$.

To illustrate the application of RTLP, the stepwise results for a clustering scenario are presented. As shown in Figure \ref{clustering_example} (a), 120 curves and four clusters with a cardinality of 30 are generated. Subsequently, random 12 out of 120 curves are replaced with outliers marked in dashed orange. Then, missing values are generated in random 96 out of 120 curves. Here, the standard time grid has 20 equidistant points from 0 to 1. For the curves with missing values, six values are missing randomly in the standard time grid on average, and the locations of the missing measurements are independent and random per subject. The transparency of the color increases with the time elapsed. 

Algorithms \ref{clustalg}-\ref{outalg} are executed under $p_m = 0.1$, $\alpha=0.85$ and $\theta \in [0, 1]$ after the calculation of the ETD. Figure \ref{clustering_example} (b) only visualizes the change of average silhouette value versus $\theta$ when $\theta \leq 0.3$ since it already reaches zero when $\theta = 0.26$. In addition, Figure \ref{clustering_example} (b) indicates an increase in $\bar{s}(\theta)$ for $\theta <0.08$ when the number of clusters decreases significantly to four; a stable $\bar{s}(\theta)$ for $0.08 \leq \theta \leq 0.21$, corresponding to the correct number of clusters with a slight movement of the outlier set; and a decrease in $\bar{s}(\theta)$, when the number of clusters decreases from four to one. The optimal $\theta$ is $0.08$, corresponding to the highest average silhouette value. Subsequently, the clustering result is that of Algorithms \ref{clustalg}-\ref{outalg} under $\theta = 0.08$. Figures \ref{clustering_example} (c) and (d) show the results of the first- and second-layer partitions, i.e., $\bm{G}$ and $\bm{C}$, respectively, from Algorithm \ref{clustalg}. The first four sets in $\bm{G}$ and $\bm{C}$ merge the majority of the curves and are successively marked in black, red, green, and blue. The remaining sets in $\bm{G}$ and $\bm{C}$ have a cardinality of either one or two with other distinct colors per set. Figures \ref{clustering_example} (e) and (f) show identical results obtained before and after updating the primary clusters $\bm{C}_p$ via Algorithm \ref{outalg}. On applying Criterion \ref{cri3}, all potential outliers remain as outliers with all true outliers correctly detected, as seen in Figure \ref{clustering_example} (f).

Algorithm \ref{distan_alg} takes $ETD(\widehat{\bm{Y}}_i, \widehat{\bm{Y}}_j)$ as the basic operation. Algorithm \ref{clustalg} takes ${G}^M \gets nbr(core(\widetilde{S}, \theta), \\ \widetilde{S}, \theta)$ as the basic operation. Algorithm~\ref{outalg} takes $ETD(\bm{X}, core(C, \theta)) > q_{\alpha}(D(C, core(C, \theta)))$ as the basic operation. The computational complexity of the worst cases in the above algorithms take $\mathcal{O}(N^2)$, $\mathcal{O}(N)$, and $\mathcal{O}(|\bm{C}_p|\times|O|)$, respectively. Here, $|\bm{C}_p|\leq I$, $|O|\leq N p_m I$ and empirically, $I \leq c\sqrt{N}$ for some $c>0$. Therefore, most of the running time is spent on Algorithm \ref{distan_alg}, and this situation can be improved by parallel computation (\citeauthor{rossini2007simple} \citeyear{rossini2007simple}).

\section{Simulation Study}
\label{sec3}
The aim of the simulation is to construct several scenarios, including the corruption from both outliers and time sparseness, and to demonstrate the robustness of the proposed algorithm in comparison with existing methods. As to the assessment indexes, the adjusted Rand index (ARI, \citeauthor{hubert1985comparing} \citeyear{hubert1985comparing}, \citeauthor{ferreira2009comparison} \citeyear{ferreira2009comparison}), percentage of correct outlier detection ($p_c$, i.e., the number of correctly detected outliers divided by the number of outliers), and percentage of false outlier detection ($p_f$, i.e., the number of falsely detected outliers divided by the number of nonoutliers) are adopted to assess the goodness of clustering and outlier detection. The settings with outliers and time sparseness are considered in Subsection \ref{sim}, and the comparison results are given in Subsection \ref{simulation}. The RTLP, the ETD-based DBSCAN, agglomerative hierarchical clustering, and $K$-medoids methods, and the model-based funHDDC (\citeauthor{schmutz2020clustering} \citeyear{schmutz2020clustering}) are examined. The optimal $K$ (or $\theta$) was determined by the average silhouette value for the ETD-based agglomerative hierarchical clustering and $K$-medoids (or ETD-based RTLP and DBSCAN) and by the BIC for funHDDC.

\subsection{Simulation Settings}
\label{sim}
Here, six types of scenarios are listed, and the contamination and the time sparseness are introduced in each scenario. The considered clustering scenarios are as follows: amplitude variation, phase variation, horizontal shift, clover petals as closed curves, cyclone tracks mimicked by the curves starting from a similar location and diverting to various destinations, and helixes with different radii. 

For the sample $\bm{f}_i~(i=1,\ldots,N)$ evaluated at the common time grid $\bm{st}=\{\frac{k-1}{T-1}, k=1, \ldots, T\}$, $\bm{f}_i=\bm{m}_i+\bm{e}_i$. For $t \in \bm{st}$, $\bm{m}(t)=(m^{(1)}(t), \ldots, m^{(p)}(t))^{\top}$, and $\bm{e}_i(t)=(e^{(1)}(t), \ldots, e^{(p)}(t))^\top$. The measurement errors evaluated at the standard grid are denoted as $\bm{e}=(\bm{e}^1, \bm{e}^2, \ldots, \bm{e}^p)^\top$, where $\bm{e}^k=\{e^{(k)}(t), t \in \bm{st}\}$. It is assumed that $\bm{e}$ follows a normal distribution $\mathcal{N}_{pT}(\bm{0}, \bm{\Sigma})$, where $\bm{\Sigma}$ is a covariance matrix consisting of blocks $\bm{\Sigma}_{i,j} \in \mathbb{R}^{T \times T}$ for $1 \leq i, j \leq p$, and $\Sigma_{i,j}(s, t)$ denotes the covariance between $e^{(i)}(s)$ and $e^{(j)}(t)$. Assume that $\Sigma_{i, j}(s, t)$ follows the Mat{\'e}rn cross-covariance (\citeauthor{gneiting2010matern} \citeyear{gneiting2010matern}) function as follows:
\begin{align*}
\Sigma_{i,j}(s, t)&=
\begin{cases}
\sigma_{i}^2{\cal M}(|s-t|; \nu_{i}, \eta_{i}), &~~~~1 \leq i=j \leq p,\\
\rho_{i,j}\sigma_{i}\sigma_{j}{\cal M}(|s-t|; \nu_{i,j}, \eta_{i,j}),& ~~~~1 \leq i\neq j \leq p, 
\end{cases}
\end{align*}
where ${{\cal M}(h;\nu,\eta)=\frac{2^{1-\nu}}{\Gamma(\nu)}(\eta h)^{\nu}\mathcal{K}_{\nu}(\eta h)}$, ${ h=|s-t|}\in [0,1]$, and $\mathcal{K}_{\nu}$ is a modified Bessel function of the second kind of order $\nu$. Here, $\sigma^2$ is the marginal variance, $\nu>0$ adjusts smoothness, and $\eta>0$ is a scale parameter. Let $\nu_{i,j}=\frac{1}{2}(\nu_{i}+\nu_{j})$, and $\nu_i$ is generated from the uniform distribution, $\nu_i\sim \mathcal{U}(0.2, 0.3)$. Let the matrix $(\beta_{i,j})^{p}_{i,j=1}$ be symmetric and nonnegative definite with diagonal elements $\beta_{i,i} = 1$ for $i = 1,\ldots, p$, and nondiagonal elements $\beta_{i,j}\sim \mathcal{U}(0, 1)$ for $1 \leq i<j\leq p$, and $\rho_{i,j} = \beta_{i,j}\frac{\Gamma(\nu_i + \frac{p}{2})^{1/2}}{\Gamma(\nu_i)^{1/2}} \frac{\Gamma(\nu_j + \frac{p}{2})^{1/2}}{\Gamma(\nu_j)^{1/2}} \frac{\Gamma\{\frac{1}{2}(\nu_i + \nu_j)\}}{\Gamma\{\frac{1}{2}(\nu_i + \nu_j)+\frac{p}{2}\}}$
for $1 \leq i \leq j \leq p$.

Let the number of variables $p=3$, and $T=50$ equidistant points in $[0,1]$; the number of objects $N = 150$; the number of clusters $K = 3$. Here, $\sigma^2_1=0.05$, $\sigma_2^2=0.2$, and $\sigma_3^2=0.3$. Each cluster has an equal number of samples. For $k= 1, \ldots, K$ and $v = 1,\ldots, p$, $l \sim \mathcal{U}(1, p)$, and $r\sim \mathcal{B}(1/2)$, where $\mathcal{B}(b)$ represents a Bernoulli distribution with probability $b$. All six clustering scenarios are described as follows:

Scenario 1 (amplitude variation): 
${m}^{(v)}(t)=2 v \cos\{(l+rv/4) \pi t\}+(-1)^r 3 k v $. 

Scenario 2 (phase variation): ${m}^{(v)}(t)=2 v \cos\{(l+rv/4) \pi h(t)\}$. Let $h(t)=\log_2(t+1)$ when $k=1$; $h(t)=t^2$ when $k=2, v = 1$; $h(t)=1-\cos(\pi t/2)$ when $k=2, v = 2$; $h(t)=\sin^2(\pi t/2)$ for $k=2, v = 3$; $h(t)=t^3$ for $k=3, v = 1$; $h(t)=\sin(\pi t/2)$ for $k=3, v = 2$; and $h(t)=t$ for $k=3, v=3$.

Scenario 3 (shift variation): 
${m}^{(v)}(t)= 6\cos\{(l+v)\pi (t + 0.2 l k)/2\}$.

Scenario 4 (clover petals): Let $m^{(1)}(t)=5 \cos \{3 w_k(t)\} \cos \{w_k(t)\}$, $m^{(2)}(t)=5 \cos \{3 w_k(t)\} \cdot \sin \{w_k(t)\}$, and $m^{(3)}(t)=5 \cos \{3 w_k(t)\}$, where $w_k(t) = 1.01(t+k-1) + 0.548$. 

Scenario 5 (cyclone tracks):~\\ 
${m}^{(1)}(t) =
\begin{cases}
-5t-\sin(10\pi t),\\
5\sin(10\pi t), \\
5t+\sin(10\pi t), \\
\end{cases}$
${m}^{(2)}(t) =
\begin{cases}
11 t^2,\\
7t,\\
5\log_2(t+1),\\
\end{cases}$
${m}^{(3)}(t) =
\begin{cases}
7\log_2(t+1), & k=1,\\
14\log_2(t+1), & k=2,\\
21\log_2(t+1), & k=3.\\
\end{cases}$

Scenario 6 (helixes with variable radii): ~\\
${m}^{(1)}(t) = \begin{cases}
5\cos(10\pi t), \\
5t \cos(20\pi t+10), \\
5t \sin(20\pi t+10), \\
\end{cases}$
${m}^{(2)}(t) = \begin{cases}
5.5\sin(10\pi t), \\
5t \sin(20\pi t+10),\\
5\log_2(t+1),\\
\end{cases}$
${m}^{(3)}(t) = \begin{cases}
10t, & k=1,\\
10 - 10t, & k=2,\\
10t, & k = 3.\\
\end{cases}$

\begin{figure}[ht]
     \begin{subfigure}[h]{\textwidth}
         \centering
         \includegraphics[height=4.5cm,width=\textwidth]{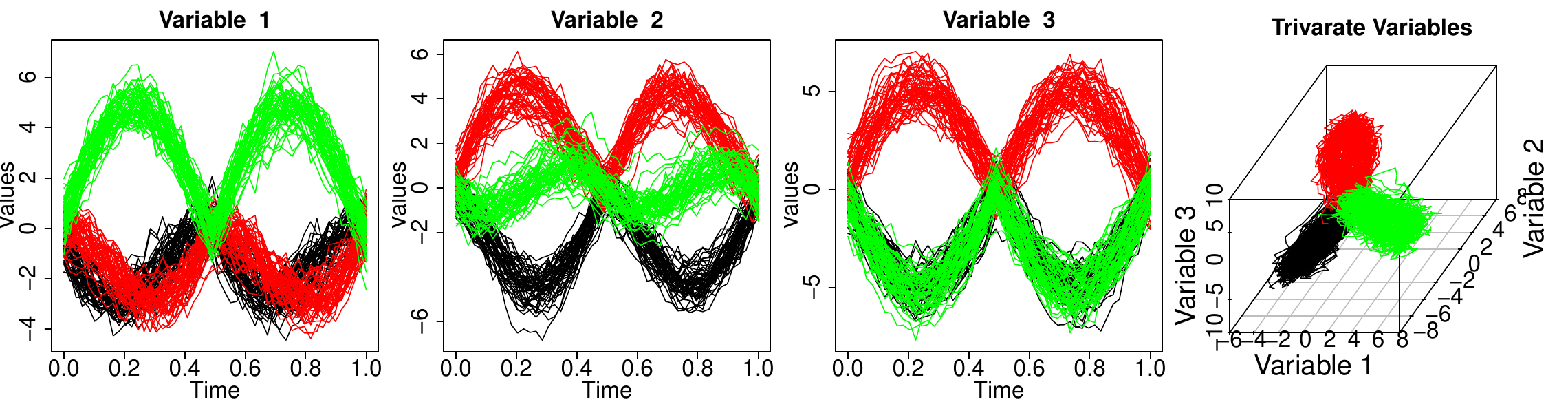}
     \end{subfigure}
     \begin{subfigure}[h]{\textwidth}
         \centering
         \includegraphics[height=4.5cm,width=\textwidth]{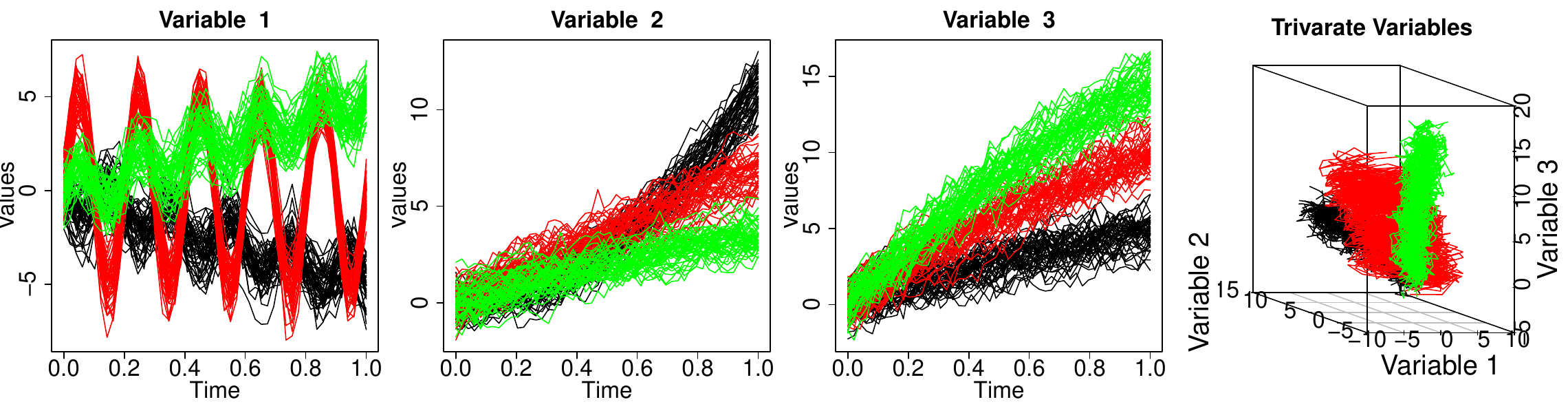}
     \end{subfigure}
\caption{Top and bottom rows represent samples from Scenarios 4 and 5, respectively. Three clusters are represented in black, red, and green.}
\label{models}
\end{figure}
Scenarios 4 and 5 are representatives of closed and nonclosed curves. The visualization (Figure \ref{models}) and the clustering results of Scenarios 4 and 5 are provided in this paper; the results for the remaining scenarios are provided in the supplementary material.

\begin{figure}[ht!]
     \begin{subfigure}[h]{\textwidth}
         \centering
         \includegraphics[height=5cm,width=15cm]{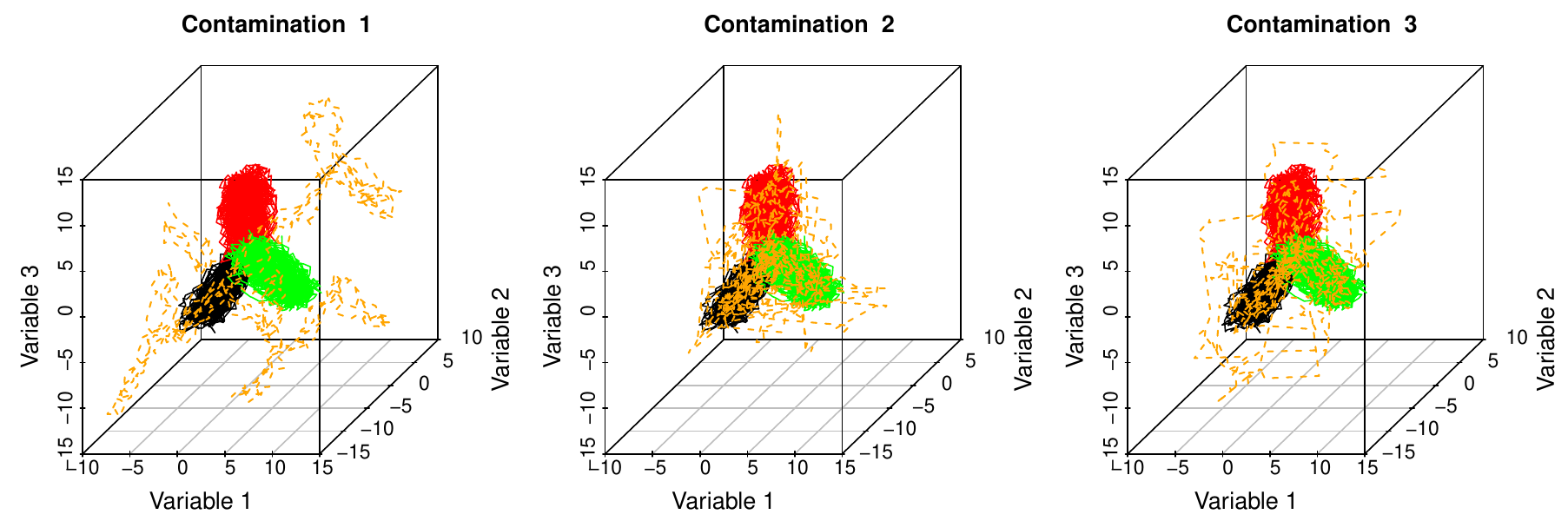}
         \includegraphics[height=5cm,width=15cm]{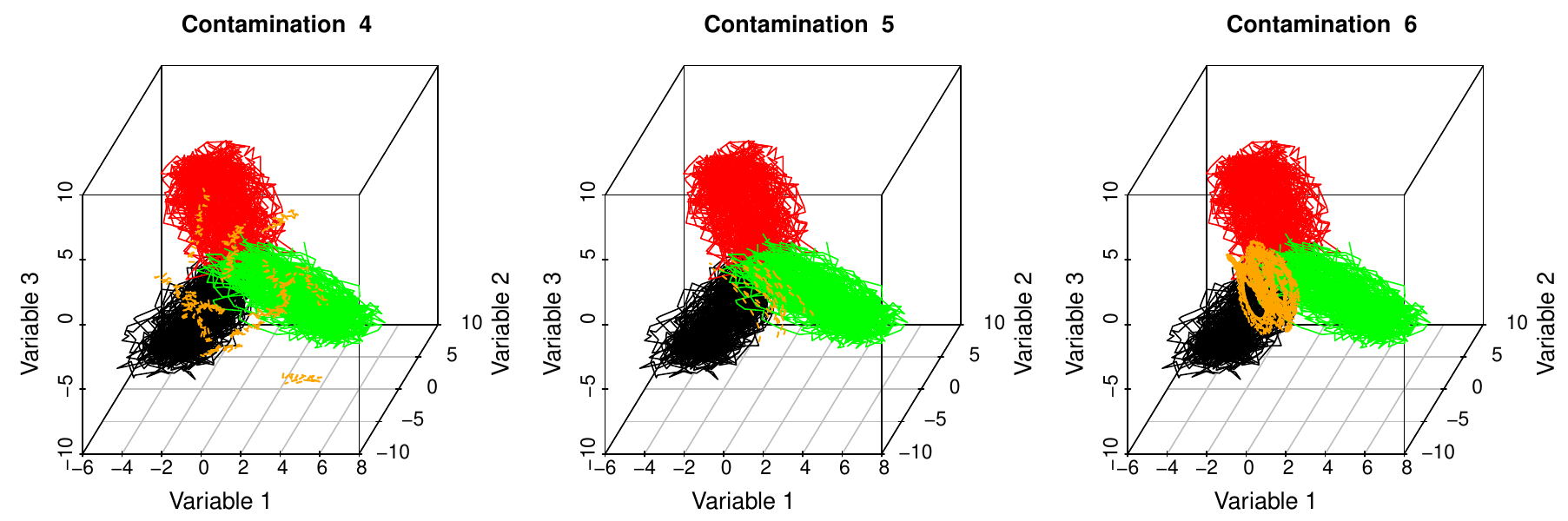}
     \end{subfigure}
\caption{Visualization of Scenario 4 with a proportion 10\% of outliers. Three clusters are labeled in solid black, red, and green, whereas the outliers are labeled in dashed orange.}
\label{outl_model}
\end{figure}

Next, three types of magnitude outliers (Contaminations 1-3), and three types of shape outliers (Contaminations 4-6) are introduced, with a 10\% proportion of samples in the aforementioned clustering scenarios.~The visualization of Scenario 4 with various outlier settings is shown in Figure \ref{outl_model}. The first and second rows in Figure \ref{outl_model} reveal the outliers showing abnormalities in magnitude and shape, respectively. It is noticed that outliers in Contaminations 1-4 are scattered and those in Contaminations 5-6 are concentrated.
~\\
Contamination 1 (pure outlier):
$\widetilde{m}^{(v)}_i(t)=m^{(v)}_i(t)+rh$. Here $r$ is assigned to 1 if $\mathcal{U}(-1,1)>0$, otherwise $r$ is $-1$, and $h = \max\{|\min(m^{(v)}_i)(t)|, |\max(m^{(v)}_i)(t)|\}/2$. Then, $r$ and $h$ below are generated in the same way as Contamination 1.
~\\
Contamination 2 (peak outlier): $\widetilde{m}^{(v)}_i(t)=
\begin{cases}
m^{(v)}_i(t)+rh, & t \in [st, st+0.1], \\
m^{(v)}_i(t), & {\rm otherwise},\\
\end{cases}$ where $st\sim \mathcal{U}(0, 0.9)$.
~\\
Contamination 3 (partial outlier): $\widetilde{m}^{(v)}_i(t)=
\begin{cases}
m^{(v)}_i(t)+rh, & t \in [st, 1], \\
m^{(v)}_i(t), & {\rm otherwise},\\
\end{cases}$ where $st\sim \mathcal{U}(0, 0.5)$.
~\\
Contamination 4 (shape outlier I):
$c \sim \mathcal{U}(\min(m^{(v)}_i)(t)/2, \max(m^{(v)}_i)(t)/2)$, $s\sim \mathcal{U}(-2v, 2v)$,
$\widetilde{m}_i^{(v)}(t)=c+s t+\epsilon_i(t)$, and $\epsilon_i(t)\sim \mathcal{U}(-0.3,0.3)$.
~\\
Contamination 5 (shape outlier II): 
$\widetilde{m}^{(1)}_i(t) = s^{(1)}(t) + h\cos(0.5\pi t)+ a$, $\widetilde{m}^{(2)}_i(t) = s^{(2)}(t) + h\sin(0.5\pi t) + a$, and $\widetilde{m}^{(3)}_i(t) = s^{(3)}(t) - h\cos(0.5\pi t) + a$, where $s^{(v)}(t)=\min(m^{(v)}_i)(t)/2 + \max(m^{(v)}_i)(t)/2~(v=1, 2, 3)$, and $a \sim \mathcal{U} (-h, 0)$. The $s^{(v)}(t)$ below is generated in the same way.
~\\
Contamination 6 (shape outlier III):
$\widetilde{m}^{(1)}_i(t) = s^{(1)}(t) + h\cos(30 \pi t)+ a$, $\widetilde{m}^{(2)}_i(t)=s^{(2)}(t) + h\sin(30 \pi t) + a$, and $\widetilde{m}^{(3)}_i(t)=s^{(3)}(t) - h\cos(30 \pi t) + a$, where $a \sim \mathcal{U} (-h, 0)$.

In addition, the parameters $p_{size}$ and $p_{curve}$ are introduced to ensure that each curve has unique time measurements in order to verify that the RTLP clustering algorithm can handle sparse multivariate functional data. Here, $p_{size}$ represents the number of curves with missing values divided by the total number of observed curves, and $p_{curve}$ represents the number of missing points for curves with missing values divided by the cardinality of the standard time grid. The simulation for $p_{size}=100\%$ and $p_{curve}$ is conducted ranging from $0\%$ up to $60\%$. 

\begin{figure}[ht!]
    \centering
    \includegraphics[height=5cm, width=15cm]{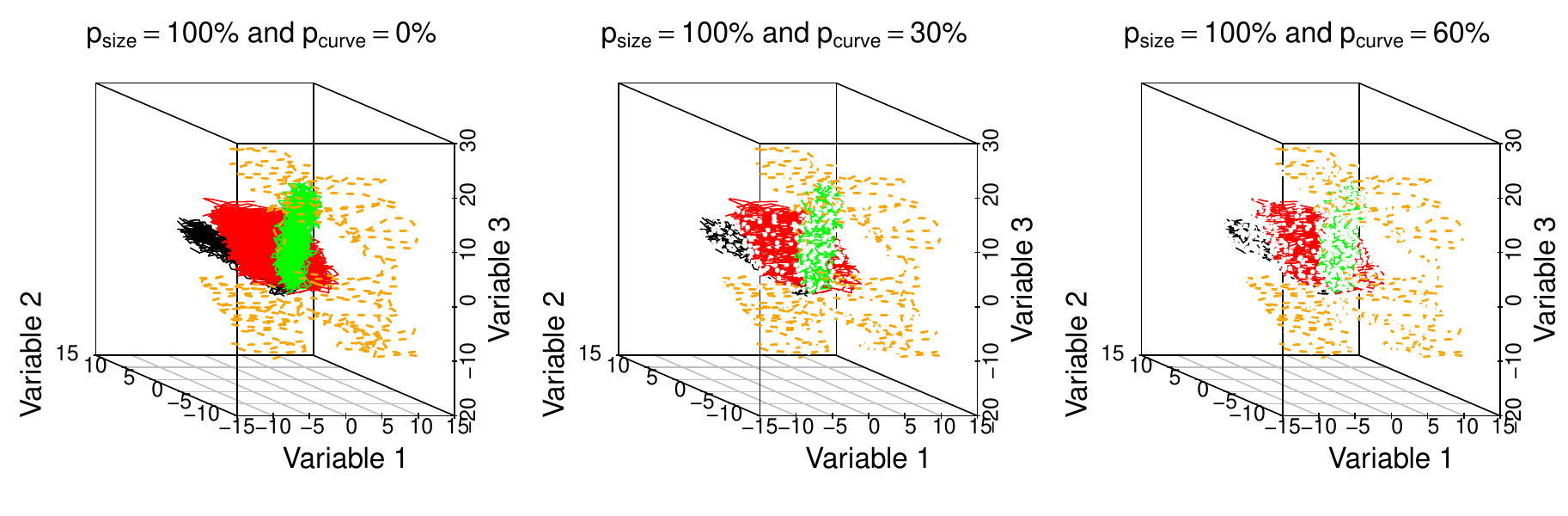}
    \caption{Visualization of Scenario 5 with 10\% pure magnitude outliers for $p_{curve} = 0\%, 30\%$, and $60\%$ separately and $p_{size}=100\%$ in all above settings. The three clusters are shown in black, red, and green, and the outliers are indicated in orange.}
    \label{sparse_visualization}
\end{figure}

Figure \ref{sparse_visualization} shows the visualization of Scenario 5 with 10\% proportion of pure magnitude outliers for different time sparseness $p_{curve} = 0\%$, $30\%$, and $60\%$ and $p_{size} = 100\%$, respectively. As data trajectories become sparser, the clustering patterns are harder to recognize, and the clustering is expected to become more challenging. 

\subsection{Clustering Performance}
\label{simulation}
 The ARI, $p_c$, and $p_f$ are applied to assess the goodness of clustering and outlier detection. ARI can be applied when the actual clusters are known. It estimates the matching between the actual partition and the estimated partitions, and it is zero in the case of random partitions and one in the case of perfect agreement between two partitions.
 \begin{figure}[b!]
    \centering
    \includegraphics[width=\textwidth,height=5.7cm]{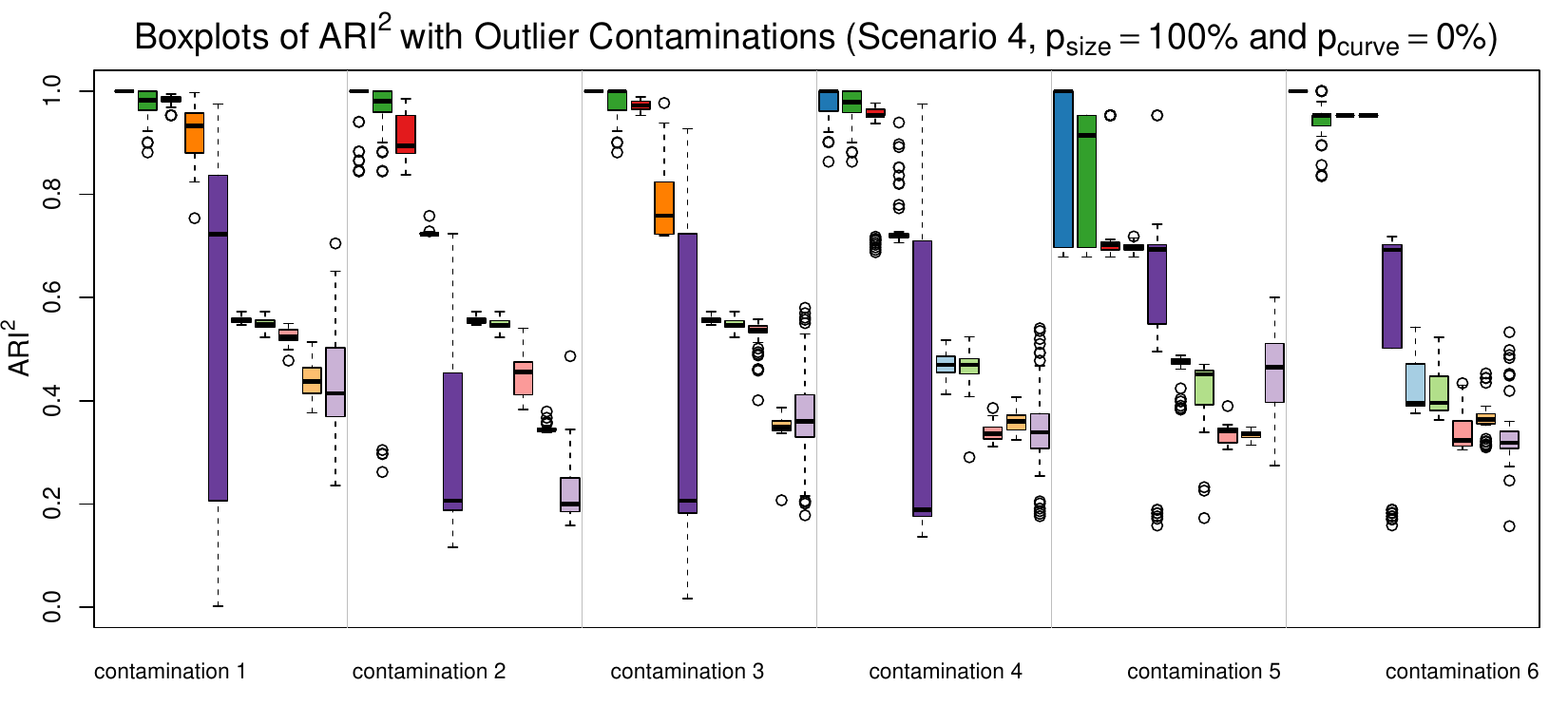}
    \includegraphics[width=\textwidth,height=5.7cm]{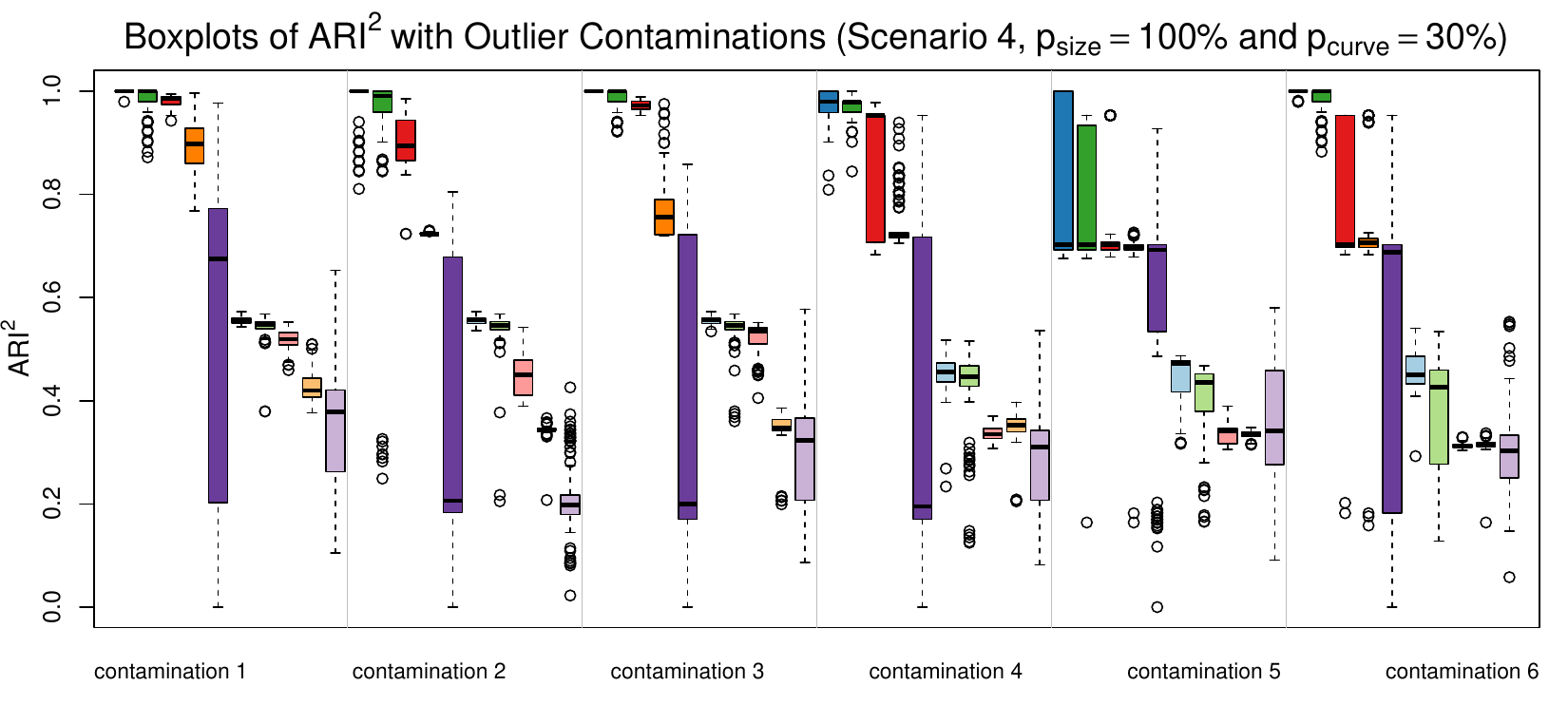}
    \includegraphics[width=\textwidth,height=6.3cm]{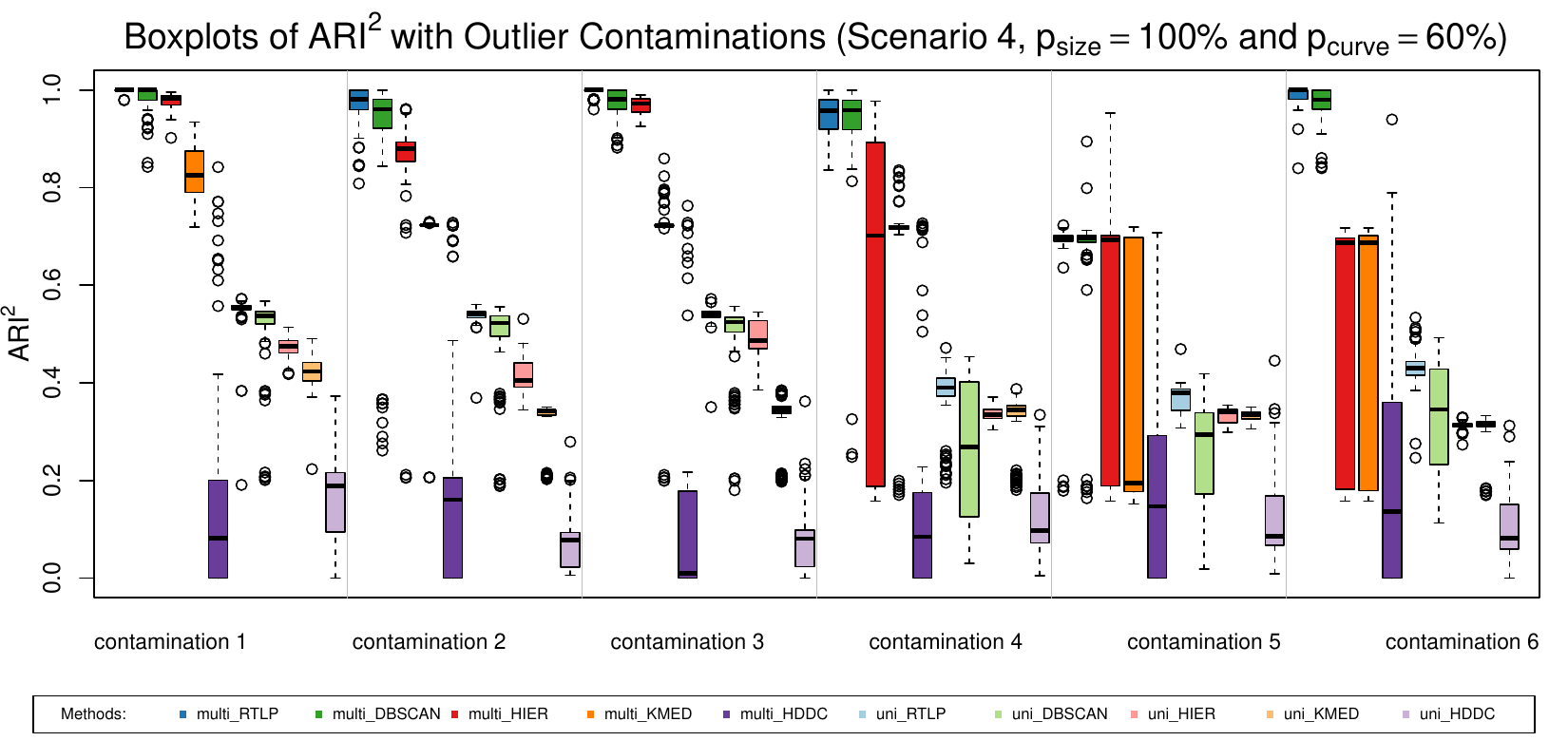}
    \caption{Panels from the top to bottom display the boxplots of ARI$^2$ in Scenario 4 for $p_{curve} = 0\%, 30\%$ and $60\%$. Ten methods are compared under all settings with six contaminations. Here, $K=3$, and there are 100 simulation replicates. The methods, from left to right, are the multivariate and average marginal univariate versions of RTLP, DBSCAN, agglomerative hierarchical clustering, $K$-medoids, and funHDDC methods.}
    \label{result_sce4}
\end{figure}

To illustrate the robustness of the proposed algorithm, ARI is compared among ETD-based RTLP, DBSCAN, agglomerative hierarchical clustering, $K$-medoids methods, and funHDDC. In particular, these five methods are applied to multivariate functional data and to each marginal functional data separately. The ARI for multivariables and the average ARI over marginal variables are obtained to evaluate the investigated method under the two situations. The method applied to multivariate (univariate) functional data is denoted as its multivariate (univariate) version. Note that the difference between ETD-based multivariate clustering methods and univariate clustering ones starts in the calculation of ETD; see step 1) in Section \ref{clustering_execution}. 

\begin{figure}[b!]
     \centering
    \includegraphics[width=\textwidth,height=5.7cm]{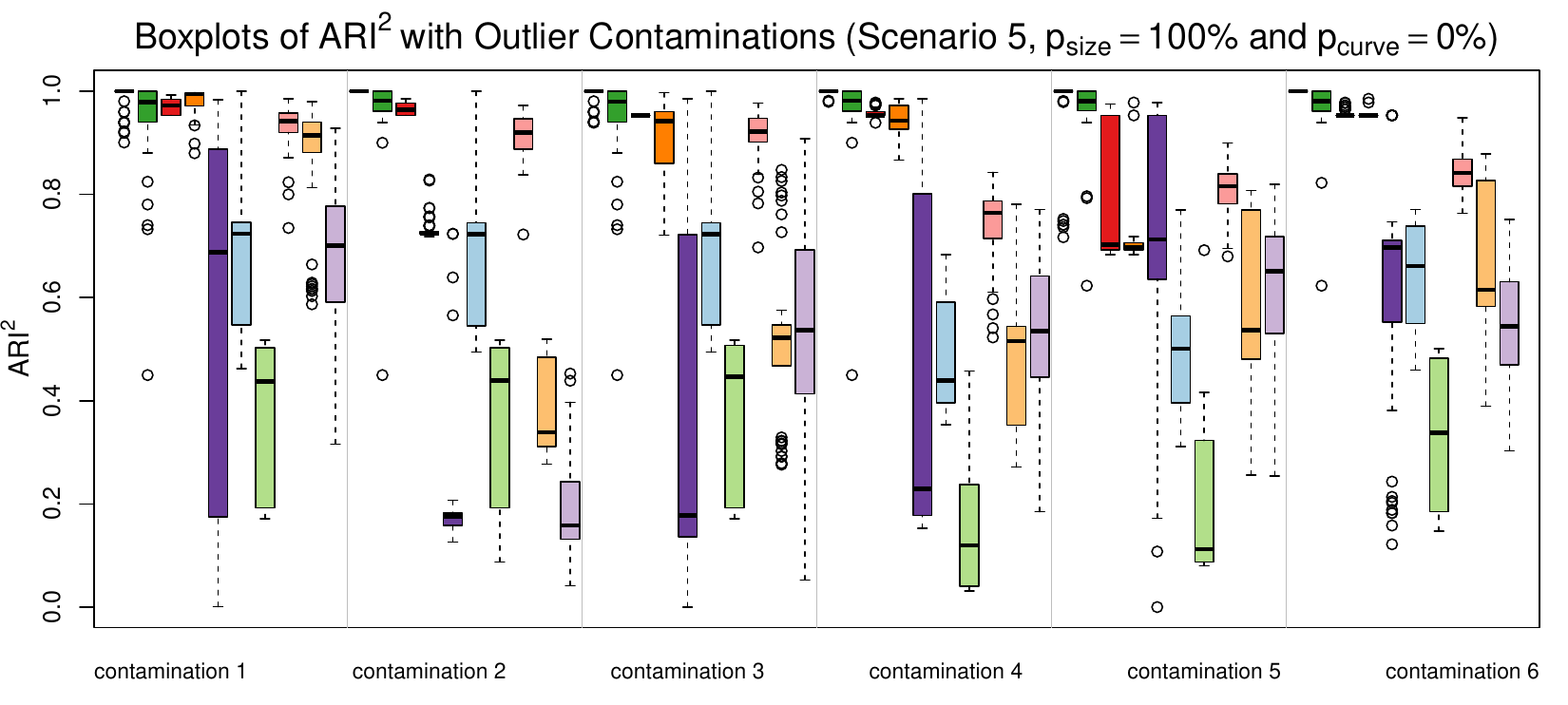}
    \includegraphics[width=\textwidth,height=5.7cm]{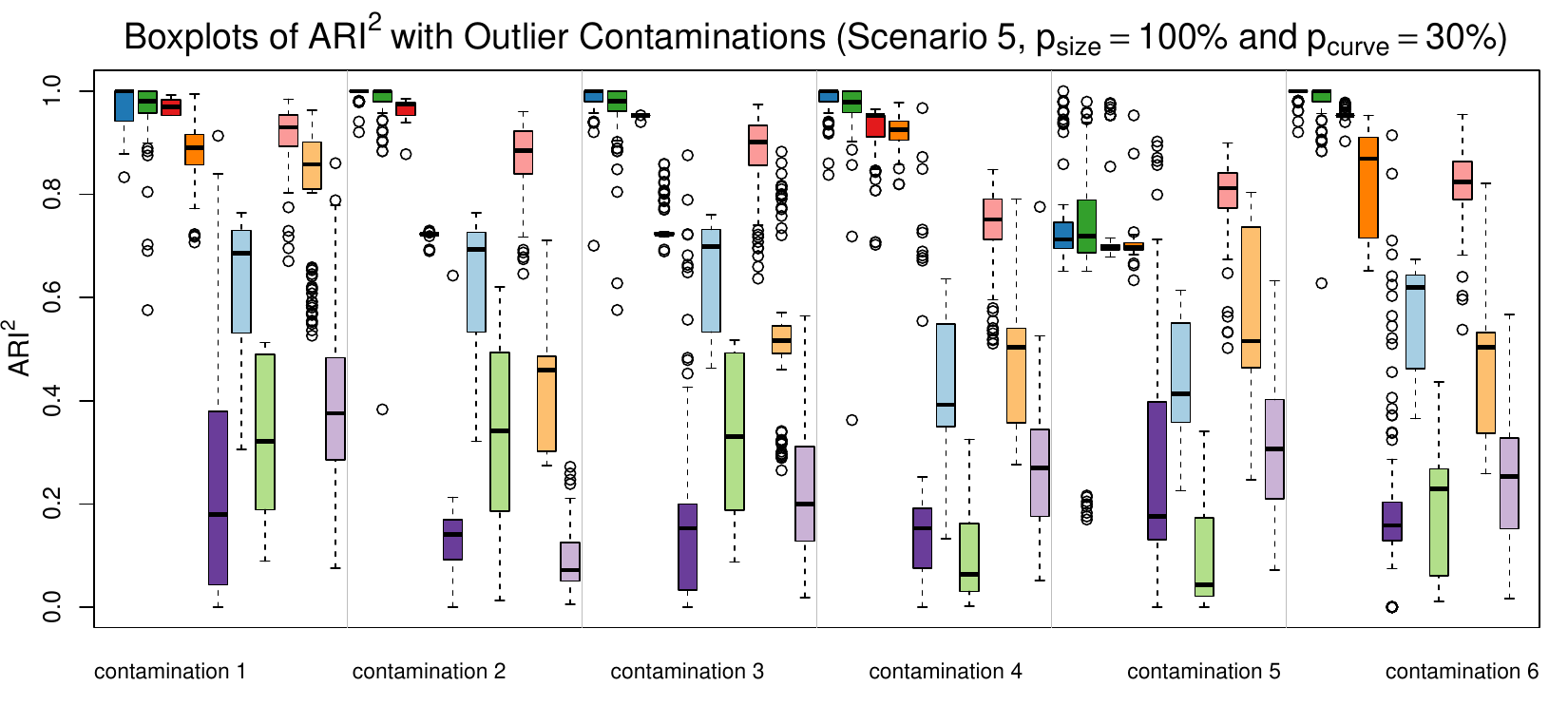}
    \includegraphics[width=\textwidth,height=6.3cm]{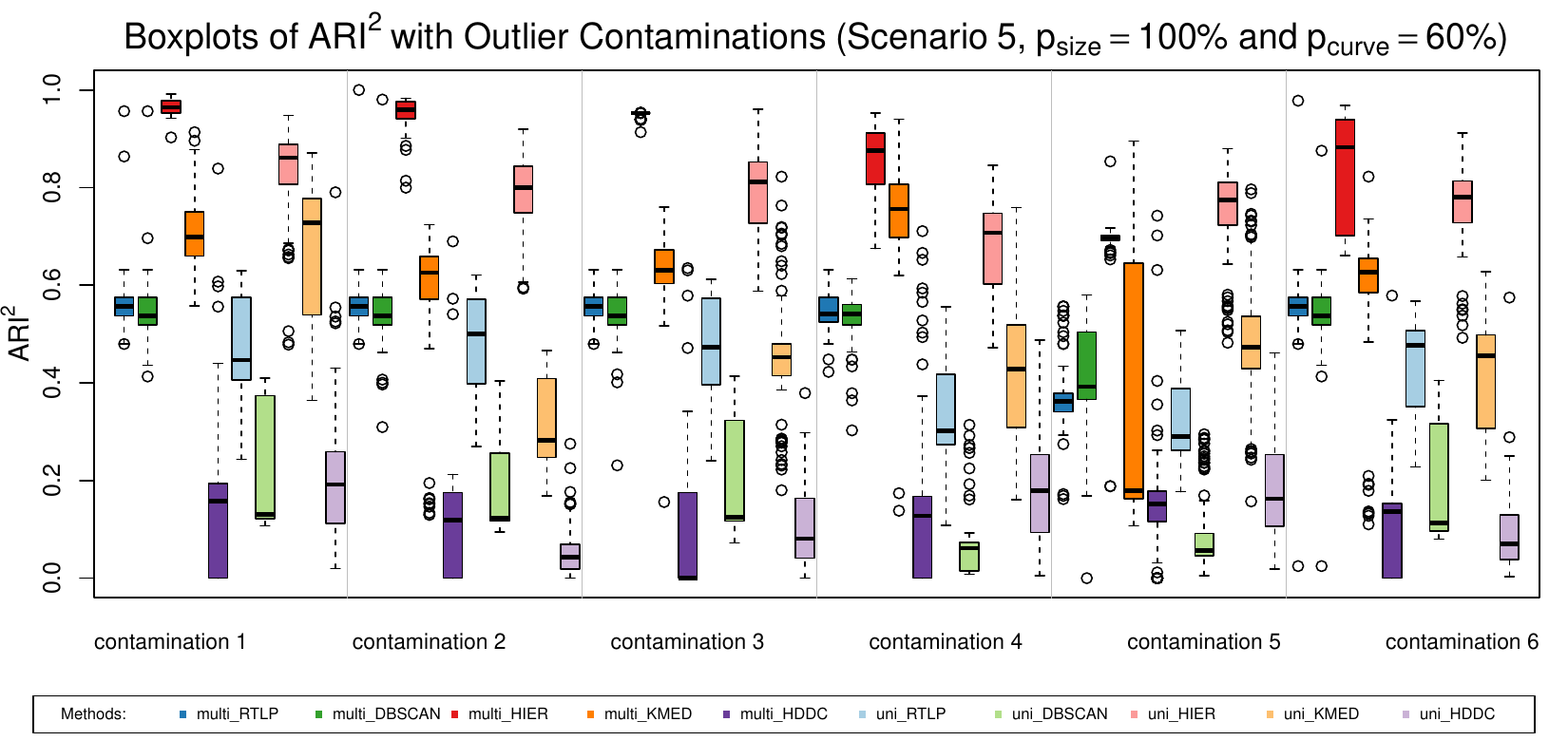}
    \caption{Panels from the top to bottom display the boxplots of ARI$^2$ in Scenario 5 for $p_{curve} = 0\%, 30\%$ and $60\%$. Ten methods are compared under all settings with six contaminations. Here, $K=3$, and there are 100 simulation replicates. The methods, from left to right, are the multivariate and average marginal univariate versions of RTLP, DBSCAN, agglomerative hierarchical, $K$-medoids, and funHDDC methods.}
    \label{result_sce5}
\end{figure}

Figures \ref{result_sce4} and \ref{result_sce5} show the boxplots of ARI$^2$ for Scenarios 4 and 5 under different outlier contaminations and missing time measurements. Overall, the ETD-based clustering methods perform better than funHDDC, and the multivariate RTLP exhibits the best performance in achieving the ARI with the highest mean and has a very low standard deviation in almost all cases. The results are viewed from horizontal and vertical perspectives.

From a horizontal perspective of Figures \ref{result_sce4} and \ref{result_sce5}, there are three summaries under the same time measurement setting. First, the multivariate ETD-based methods perform better than their corresponding univariate marginal methods in the clustering, as seen from the median of the boxplots. Second, the ETD-based clustering methods have better clustering precision than funHDDC, as seen from the differences among the boxplots in terms of the median and range. Third, outlier contaminations may lead to a drop in the clustering precision depending on the outlier types. For example, Contamination 5 in Scenarios 4 and 5 influence the clustering precision of all ETD-based methods, even when $p_{curve} = 0\%$. In most cases, RTLP is rarely influenced except for Contamination 5 in Scenario 4 and Scenario~5 with $p_{curve}=60\%$.

From a vertical perspective of Figures \ref{result_sce4} and \ref{result_sce5}, two patterns are observed. First, the clustering pattern influences the clustering precision of the ETD-based methods when there are missing values. The precision of clustering in Scenario 4 remains almost identical when $p_{curve}$ changes from $0\%$ to $30\%$ and then to $60\%$, except for Contamination 5; the multivariate RTLP exhibits the best performance, followed by multivariate DBSCAN, multivariate agglomerative hierarchical clustering, and multivariate $K$-medoids. However, the clustering precision of multivariate RTLP and multivariate DBSCAN in most contaminations in Scenario 5 with $p_{curve}=60\%$ is not as good as multivariate agglomerative hierarchical clustering and multivariate $K$-medoids. From the pattern of Scenario 5 (Figure \ref{sparse_visualization}), it is speculated that because of too much sparseness, ETD between a curve in the middle cluster and a curve at the edge of the remaining clusters gets small and the multivariate RTLP cannot separate curves into the correct three clusters regardless of $\theta$. However, the multivariate agglomerative hierarchical clustering seeks to build a hierarchy of clusters from each observation as a separate cluster to all observations merged into one cluster, and the multivariate $K$-medoids considers the case of three clusters before determining the optimal number of clusters. Second, multivariate and univariate funHDDC methods are easily affected by outliers when there are no missing values, and their clustering performances are worse when there are missing values, which can be seen from the movement of the boxplots vertically.

The multivariate RTLP usually behaves the best when the curve sparseness $p_s$ is not greater than $50\%-60\%$. Although the results are not reported in this paper, the running time is stable for all ETD-based clustering methods but shows a significant standard deviation for funHDDC. Additionally, the multivariate RTLP achieves the best clustering performance under scenarios with amplitude and phase variation. Although it works well for the considered misaligned data when phase/amplitude variations are present (see supplementary material), more complex misaligned data structures are deferred to future research to fully resolve the challenge of sparse multivariate functional data clustering specifically for misaligned data. It is currently believed that the RTLP method can deal with mildly  misaligned data. Hence, multivariate RTLP is recommended when the data objects, i.e., multivariate functional data, are not too sparse, and it is acknowledged that the ETD-based multivariate DBSCAN and multivariate agglomerative hierarchical clustering are good alternatives for very sparse cases.

\subsection{Outlier Detection Performance}
With regard to outlier detection, $p_c$ and $p_f$ are reported only for multivariate and marginal RTLP and DBSCAN methods because of their capabilities of recognizing outliers. Specifically, $p_c$ ($p_f$) is obtained from the marginal RTLP (DBSCAN) for each variable and the average $p_c$ ($p_f$) over different variables is used to represent the performance of univariate RTLP (DBSCAN) in the outlier detection.
\begin{table}[h!]
\centering
\caption{Correct outlier detection percentage $p_c$ (\%) and false outlier detection percentage $p_f$ (\%) for multivariate and univariate RTLP and DBSCAN with different outlier contaminations for Scenario 4. Higher $p_c$ and lower $p_f$ values for each setting given the contamination and $p_{curve}$ are indicated in bold.~The proportion of outliers is 10\%, and $p_{size}$ is $100\%$ for all settings. Simulations were conducted with 100 replicates.}
\label{sce4}
\scriptsize
\begin{tabular}{lccccccc}
\hline 
\multicolumn{2}{c}{\multirow{2}{*}{\diagbox{$p_{curve}$}{Methods~~~~~~~}}} & \multicolumn{2}{c}{Contamination 1}  & \multicolumn{2}{c}{Contamination 2} & \multicolumn{2}{c}{Contamination 3}\\
\cline{3-8}
 \multicolumn{2}{c}{} & {$p_c$}  & {$p_f$} & {$p_c$}  &  {$p_f$} & {$p_c$}  &  {$p_f$}\\
\hline\hline
\multirow{4}{*}{$~0\%$} & multi\_RTLP  & \bm{$100.0$} ($0.0$) & \bm{$0.0$} (0.0) & $92.4$ (17.1) & \bm{$0.0$} (0.0) & \bm{$100.0$} (0.0) & \bm{$0.0$} (0.0) \\
\cline{2-8}
\multirow{4}{*}{} & multi\_DBSCAN  & \bm{$100.0$} ($0.0$) & $0.8$ (1.1) & 92.4 (17.1) & $0.6$ (1.0) & \bm{$100.0$} (0.0) & $0.8$ (1.1) \\
\cline{2-8}
 \multirow{4}{*}{}& ~~~uni\_RTLP & \bm{$100.0$} (0.0) & 0.0 (0.1) & \bm{$99.9$} (0.4) & 0.0 (0.1) & \bm{$100.0$} (0.0) & 0.0 (0.1) \\
 \cline{2-8}
 \multirow{4}{*}{}& ~~~uni\_DBSCAN & \bm{$100.0$} (0.0) & 0.5 (0.4) & \bm{$99.9$} (0.4) & 0.5 (0.4) & \bm{$100.0$} (0.0) & 0.5 (0.4) \\
\hline
\hline
\multirow{4}{*}{30\%}& multi\_RTLP & \bm{$100.0$} (0.0) & \bm{$0.0$} (0.1) & $93.8$ (15.5) & \bm{$0.0$} (0.0) & \bm{$100.0$} (0.0) & \bm{$0.0$} (0.0) \\
\cline{2-8}
\multirow{4}{*}{}& multi\_DBSCAN & \bm{$100.0$} (0.0) & $0.7$ (1.1) & $93.8$ (15.5) & $0.5$ (0.9) & \bm{$100.0$} (0.0) & $0.5$ (0.7) \\
\cline{2-8}
\multirow{4}{*}{}& ~~~uni\_RTLP & \bm{$100.0$} (0.0) & \bm{$0.0$} (0.1) & 99.5 (1.2) & 0.0 (0.1) & 99.4 (1.3) & 0.0 (0.1) \\
\cline{2-8}
\multirow{4}{*}{}& ~~~uni\_DBSCAN & \bm{$100.0$} (0.0) & 0.6 (0.6) & \bm{$99.5$} (1.1) & 0.6 (0.6) & 99.4 (1.3) & 0.6 (0.7) \\
\hline
\hline
\multirow{4}{*}{60\%}& multi\_RTLP & \bm{$100.0$} (0.0) & \bm{$0.0$} (0.1) & $89.9$ (14.4) & \bm{$0.0$} (0.1) & $99.3$ (2.0) & \bm{$0.0$} (0.1) \\
\cline{2-8}
\multirow{4}{*}{}& multi\_DBSCAN & \bm{$100.0$} (0.0) & $0.7$ (1.1) & $89.9$ (14.5) & $0.5$ (1.1) & \bm{$99.4$} (1.9) & $0.8$ (1.1) \\
\cline{2-8}
\multirow{4}{*}{}& ~~~uni\_RTLP & 99.8 (1.6) & 0.1 (0.2) & 94.4 (3.2) & 0.1 (0.2) & 94.4 (3.4) & 0.1 (0.2) \\
\cline{2-8}
\multirow{4}{*}{}& uni\_DBSCAN & $99.9$ (0.7) & $1.1$ (1.1) & \bm{$94.5$} (3.2) & $0.9$ (0.9) & $94.9$ (3.2) & $1.0$ (1.0) \\
\hline
\hline
\multicolumn{2}{c}{\multirow{2}{*}{\diagbox{$p_{curve}$}{Methods~~~~~~~}}} & \multicolumn{2}{c}{Contamination 4}  & \multicolumn{2}{c}{Contamination 5} & \multicolumn{2}{c}{Contamination 6}\\
\cline{3-8}
 \multicolumn{2}{c}{} &  {$p_c$}  &  {$p_f$} & {$p_c$}  &  {$p_f$} & {$p_c$}  &  {$p_f$}\\
\hline\hline
\multirow{4}{*}{~0\%}& multi\_RTLP & \bm{$92.5$} (11.3) & \bm{$0.0$} (0.0) & \bm{$63.3$} (48.7) & \bm{$0.0$} (0.0) &\bm{$100.0$} (0.0) & \bm{$0.0$} (0.0) \\
\cline{2-8}
\multirow{4}{*}{}& multi\_DBSCAN & 93.3 (10.6) & $0.4$ (1.0) &  0.0 (0.0) & $0.4$ (0.7) &18.4 (39.1) & $0.8$ (1.1) \\
\cline{2-8}
\multirow{4}{*}{}& ~~~uni\_RTLP~ & 65.5 (11.2) & 0.1 (0.1) & 61.8 (11.8) & 0.1 (0.1) & 44.0 (16.3) & 0.0 (0.1) \\
\cline{2-8}
\multirow{4}{*}{}& ~~~uni\_DBSCAN~ & 67.9 (9.0) & 0.4 (0.4) & 4.1 (13.0) & 0.5 (0.5) & 7.8 (14.4) & 0.2 (0.2) \\
\hline
\hline
\multirow{4}{*}{30\%}& multi\_RTLP & $91.3$ (11.5) & \bm{$0.0$} (0.0) & 30.0 (46.1) & \bm{$0.0$} (0.1) & \bm{$99.9$} (0.9) & \bm{$0.0$} (0.1) \\
\cline{2-8}
\multirow{4}{*}{}& multi\_DBSCAN & \bm{$92.9$} (8.6) & $0.2$ (0.5) & 0.0 (0.0) & $0.2$ (0.6) & \bm{$99.9$} (0.9) & $0.6$ (1.0) \\
\cline{2-8}
\multirow{4}{*}{}&~~~uni\_RTLP~ & 59.1 (12.4) & 0.1 (0.2) & \bm{$56.3$} (17.4) & 0.1 (0.2) & 62.6 (12.4) &$0.1$ (0.1) \\
\cline{2-8}
\multirow{4}{*}{}&~~~uni\_DBSCAN~ & 61.4 (9.6) & 0.4 (0.6) & 3.0 (9.6) & 0.7 (0.7) & 64.8 (12.8) & 0.4 (0.4) \\
\hline
\hline
\multirow{4}{*}{60\%}&multi\_RTLP~ & \bm{$82.3$} (14.5) & \bm{$0.0$} (0.2) & 0.0 (0.0) & \bm{$0.1$} (0.3) & $97.3$ (6.8) & \bm{$0.0$} (0.1) \\
\cline{2-8}
\multirow{4}{*}{}&multi\_DBSCAN~ & $82.3$ (16.1) & $0.2$ (0.5) & $0.0$ (0.0) & $0.3$ (0.8) & \bm{$97.1$} (7.2) & $0.7$ (1.2) \\
\cline{2-8}
\multirow{4}{*}{}&~~~uni\_RTLP~& 31.6 (13.2) & 0.2 (0.3) & \bm{$22.3$} (16.1) & 0.3 (0.3) & 54.3 (11.3) & 0.2 (0.2) \\
\cline{2-8}
\multirow{4}{*}{}&~~~uni\_DBSCAN~& 37.4 (11.3) & 0.6 (0.7) & 0.3 (3.3) & 1.3 (1.5) & 53.7 (9.4) & 0.8 (1.3) \\
\hline
\end{tabular}
\end{table}

\begin{table}[h!]
\centering
\caption{Correct outlier detection percentage $p_c$ (\%) and false outlier detection percentage $p_f$ (\%) for multivariate and univariate RTLP and DBSCAN with different outlier contaminations for Scenario 5. Higher $p_c$ and lower $p_f$ values for each setting given the contamination and $p_{curve}$ are indicated in bold.~The proportion of outliers is 10\%, and $p_{size}$ is $100\%$ for all settings. Simulations were conducted with 100 replicates.}
\label{sce5}
\vspace{0.4cm}
\scriptsize
\begin{tabular}{lccccccc}
\hline 
\multicolumn{2}{c}{\multirow{2}{*}{\diagbox{$p_{curve}$}{Methods~~~~~~~}}} & \multicolumn{2}{c}{Contamination 1}  & \multicolumn{2}{c}{Contamination 2} & \multicolumn{2}{c}{Contamination 3}\\
\cline{3-8}
 \multicolumn{2}{c}{} &  {$p_c$}  &  {$p_f$} & {$p_c$}  &  {$p_f$} & {$p_c$}  &  {$p_f$}\\
\hline\hline
\multirow{4}{*}{~0\%}&multi\_RTLP~ & 97.8 (5.0) & $0.2$ (0.5) & \bm{$100.0$} (0.0) & \bm{$0.0$} (0.0) & $98.9$ (2.5) & \bm{$0.2$} (0.4) \\
\cline{2-8}
\multirow{4}{*}{}&multi\_DBSCAN~ & \bm{$100.0$} (0.0) & $2.2$ (4.2) & \bm{$100.0$} (0.0) & $1.3$ (3.6) & \bm{$100.0$} (0.0) & $2.2$ (4.2) \\
\cline{2-8}
\multirow{4}{*}{}&~~~uni\_RTLP & 99.9 (1.6) & \bm{$0.2$} (0.4) & 99.3 (3.1) & 0.2 (0.3) & \bm{$100.0$} (0.0) & \bm{$0.2$} (0.4) \\
\cline{2-8}
\multirow{4}{*}{}&~~~uni\_DBSCAN & \bm{$100.0$} (0.0) & 0.4 (0.6) & 99.3 (3.1) & 0.4 (0.5) & \bm{$100.0$} (0.0) & 0.4 (0.6) \\
\hline
\hline
\multirow{4}{*}{30\%} &multi\_RTLP~ & 96.4 (6.7) & \bm{$0.5$} (0.8) & \bm{$99.9$} (0.9) & \bm{$0.1$} (0.4) & $98.7$ (2.7) & \bm{$0.4$} (1.4) \\
\cline{2-8}
\multirow{4}{*}{} &multi\_DBSCAN~ & \bm{$100.0$} (0.0) & $1.6$ (3.9) & \bm{$99.9$} (0.9) & $0.6$ (1.0) & \bm{$100.0$} (0.0) & $1.5$ (4.0) \\
\cline{2-8}
\multirow{4}{*}{}& ~~~uni\_RTLP~ & 97.8 (4.8) & 0.8 (2.0) & $95.8$ (5.3) & 0.5 (1.4) & 98.0 (2.4) & 0.5 (1.4) \\
\cline{2-8}
\multirow{4}{*}{}& ~~~uni\_DBSCAN~ & 99.6 (1.2) & 1.2 (2.2) & $96.1$ (5.3) & $0.7$ (1.3) & 98.2 (2.4) & 0.6 (1.2) \\
\hline
\hline
\multirow{4}{*}{60\%} & multi\_RTLP~ & 99.8 (2.0) & 32.7 (4.5) & \bm{$100.0$} (0.0) & 33.0 (3.6) & \bm{$99.8$} (1.1) & 33.3 (1.2) \\
\cline{2-8}
\multirow{4}{*}{} & multi\_DBSCAN~ & \bm{$100.0$} (0.0) & $33.5$ (4.1) & \bm{$100.0$} (0.0) & 33.8 (3.9) & \bm{$99.8$} (1.1) & $34.1$ (2.2) \\
\cline{2-8}
\multirow{4}{*}{}& ~~~uni\_RTLP~ & 98.5 (4.0) & \bm{$11.5$} (0.9) & 95.4 (3.5) & $11.4$ (1.0) & 96.1 (3.5) & \bm{$11.3$} (1.4) \\
\cline{2-8}
\multirow{4}{*}{}& ~~~uni\_DBSCAN~ & \bm{$100.0$} (0.2) & 11.6 (0.9) & 95.5 (3.4) & \bm{$11.4$} (0.6) & 96.2 (3.5) & 11.4 (1.2) \\
\hline
\hline
\multicolumn{2}{c}{\multirow{2}{*}{\diagbox{$p_{curve}$}{Methods~~~~~~~}}} & \multicolumn{2}{c}{Contamination 4}  & \multicolumn{2}{c}{Contamination 5} & \multicolumn{2}{c}{Contamination 6}\\
\cline{3-8}
 \multicolumn{2}{c}{} & {$p_c$}  & {$p_f$} & {$p_c$}  &  {$p_f$} & {$p_c$}  &  {$p_f$}\\
\hline\hline
\multirow{4}{*}{~0\%}& multi\_RTLP~ & \bm{$99.7$} (1.3) & \bm{$0.0$} (0.0) & 89.1 (28.7) & \bm{$0.0$} (0.0) & \bm{$100.0$} (0.0) & \bm{$0.0$} (0.0) \\
\cline{2-8}
\multirow{4}{*}{}& multi\_DBSCAN~ & \bm{$99.7$} (1.3) & $1.2$ (3.6) & \bm{$95.6$} (14.4) & 0.7 (2.3) & \bm{$100.0$} (0.0) & $1.1$ (2.4) \\
\cline{2-8}
\multirow{4}{*}{}&~~~uni\_RTLP~ & $37.5$ (14.5) & $0.2$ (0.5) & 38.2 (19.9) & 0.3 (0.4) & 91.7 (14.1) & $0.2$ (0.4) \\
\cline{2-8}
\multirow{4}{*}{}&~~~uni\_DBSCAN~ & 44.3 (13.3) & 0.1 (0.2) & 19.3 (14.2) & 0.4 (1.0) & 68.7 (8.1) & 0.5 (0.7) \\
\hline
\hline
\multirow{4}{*}{30\%}&multi\_RTLP~ & 97.7 (5.4) & \bm{$0.3$} (0.8) & 21.0 (34.7) & \bm{$0.4$} (0.7) & \bm{$100.0$} (0.0) & \bm{$0.2$} (0.5) \\
\cline{2-8}
\multirow{4}{*}{}&multi\_DBSCAN~ & \bm{$97.9$} (4.7) & $1.0$ (1.5) & 29.1 (35.4) & 1.1 (1.8) & \bm{$100.0$} (0.0) & $0.9$ (2.3) \\
\cline{2-8}
\multirow{4}{*}{}&~~~uni\_RTLP~ & 36.0 (11.4) & 4.1 (4.9) & \bm{$31.4$} (14.7) & 2.2 (3.7) & 66.8 (2.3) & 0.6 (0.9) \\
\cline{2-8}
\multirow{4}{*}{}&~~~uni\_DBSCAN~ & 39.2 (10.7) & 4.8 (4.8) & 21.3 (12.2) & 3.0 (3.9) & 66.7 (0.0) & 1.3 (1.7) \\
\hline
\hline
\multirow{4}{*}{60\%}&multi\_RTLP~ & 96.3 (9.0) & 33.3 (1.2) & 19.2 (31.3) & 32.3 (5.4) & \bm{$100.0$} (0.0) & 32.6 (4.8) \\
\cline{2-8}
\multirow{4}{*}{}&multi\_DBSCAN~ & \bm{$97.9$} (4.4) & 33.9 (2.2) & \bm{$40.4$} (36.7) & $32.4$ (5.5) & \bm{$100.0$} (0.0) & $33.4$ (4.8) \\
\cline{2-8}
\multirow{4}{*}{}& ~~~uni\_RTLP~ & 33.8 (12.7) & 11.4 (0.7)& 34.0 (16.2) & 11.7 (0.8) & 71.3 (7.6) & \bm{$11.5$} (0.7) \\
\cline{2-8}
\multirow{4}{*}{}&uni\_DBSCAN~ & $37.9$ (11.4) & \bm{$11.2$} (0.5) & $22.6$ (12.8) & \bm{$11.6$} (1.0) & $76.2$ (8.2) & $11.5$ (0.9) \\
\hline
\end{tabular}
\end{table}

Tables \ref{sce4} and \ref{sce5} list $p_c$ and $p_f$ in Scenarios 4 and 5 when $p_{curve}$ increases from $0\%$ to $30\%$ and 60\%. Three phenomena are worthy to point out. First, the multivariate DBSCAN usually exhibits the highest true outlier detection percentage, and the multivariate RTLP exhibits the lowest false outlier detection percentage. This is due to the difference between DBSCAN and RTLP. DBSCAN first regards all noncore observations as potential outliers. Then, some observations from potential outliers are labeled as boundary observations of clusters during the clustering process, and the remainings are outliers. Comparatively, RTLP did not obtain potential outliers after obtaining primary clusters. Then, RTLP used a criterion to check whether potential outliers remain as outliers. Hence, the probability of a curve to be falsely (correctly) detected as an outlier is higher in DBSCAN. Second, when shape outliers are contaminated, the multivariate RTLP and DBSCAN have much better outlier detection performance compared to their marginal method, with a significantly higher $p_c$ and lower $p_f$. Third, multivariate DBSCAN achieves $p_c=0$ in Scenario 4 with Contamination 5 because it regards all outliers as one cluster instead of the outliers. Although DBSCAN and RTLP share $p_m$, DBSCAN uses $p_m$ to determine core observations, while RTLP uses $p_m$ to determine the primary clusters. From the empirical simulation, the RTLP is more robust in outlier detection compared to ETD-based DBSCAN under common $\theta$ and $p_m$.

Given the excellent performance of RTLP in the outlier-resistant clustering and outlier detection, and stable running time regardless of outlier settings and irregular time settings (running time is negatively proportional to $p_{curve}$), RTLP is highly recommended as an outlier-resistant clustering method for multivariate functional data when the average $p_{curve}$ is small to moderate. Furthermore, RTLP can effectively block outliers in the primary clusters if there are concerns about outliers. 

\section{Application to Northwest Pacific Tropical Cyclone Tracks}
\label{sec4}
\begin{figure}[ht!]
\centering
\includegraphics[width = 0.65\linewidth, height = 5.8cm]{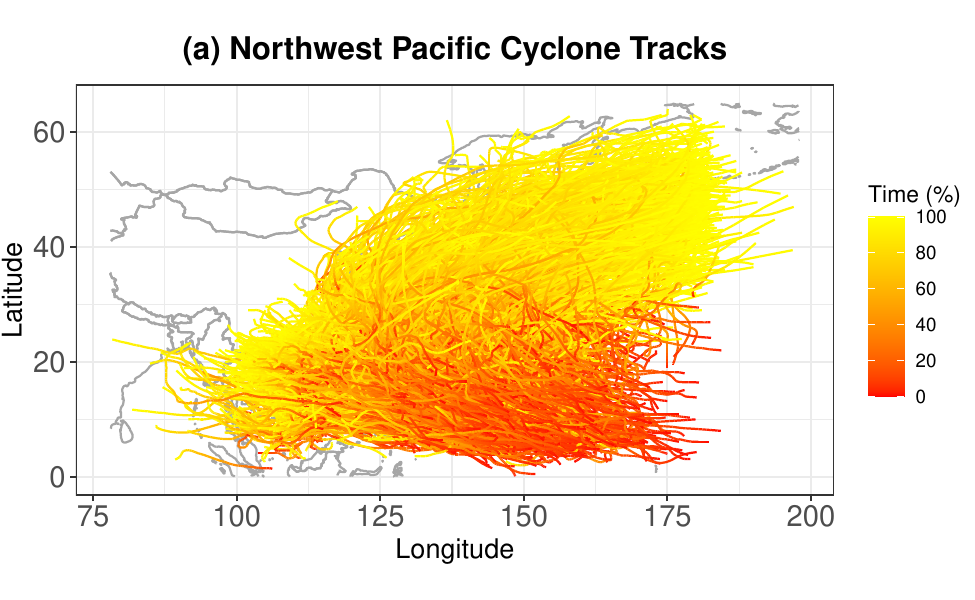}
   \begin{subfigure}{0.6\textwidth}
     \centering
     \includegraphics[width = 0.9\linewidth, height = 5.8cm]{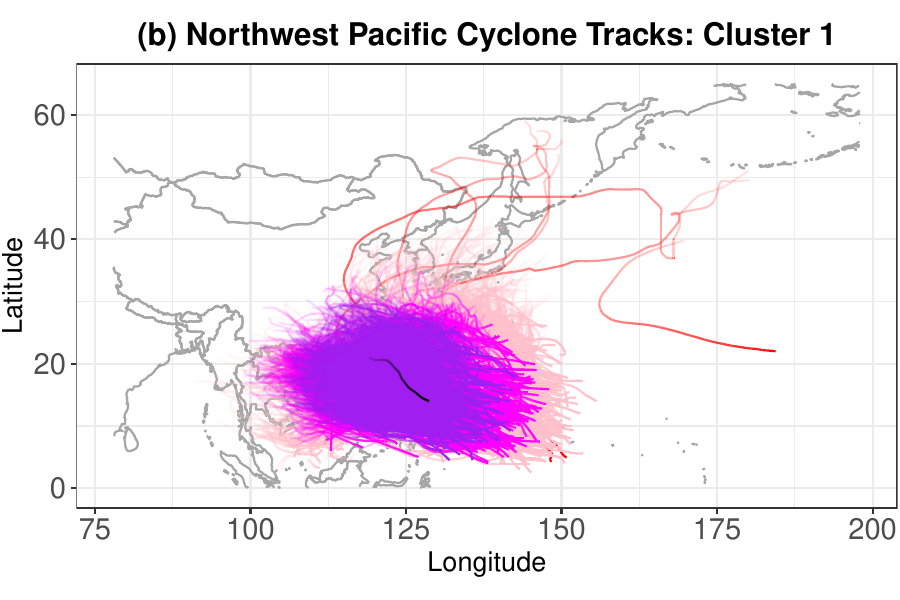}
   \end{subfigure}\hfill
   \begin{subfigure}{0.4\textwidth}
     \centering
     \includegraphics[width = 1\linewidth, height = 5.8cm]{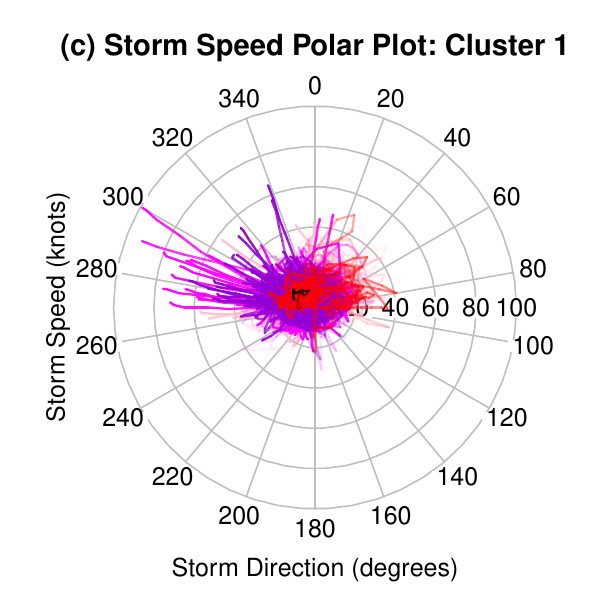}
   \end{subfigure}
   \begin{subfigure}{0.6\textwidth}
     \centering
     \includegraphics[width=0.9\linewidth, height = 5.8cm]{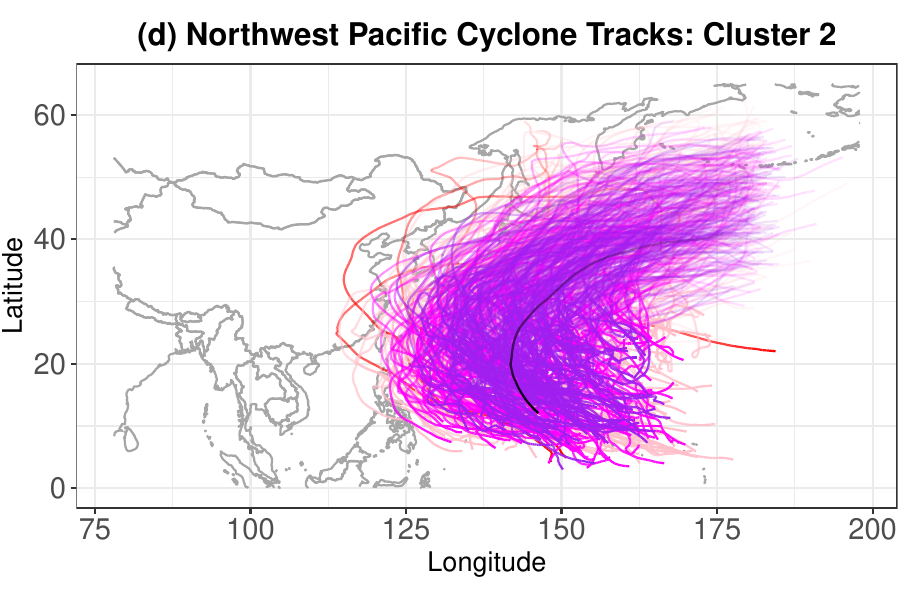}
   \end{subfigure}\hfill
   \begin{subfigure}{0.4\textwidth}
     \centering
     \includegraphics[width=1\linewidth, height = 5.8cm]{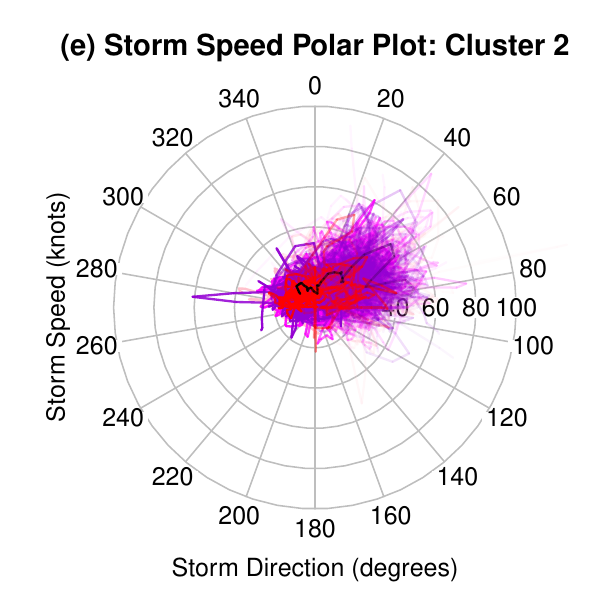}
   \end{subfigure}
   \caption{(a) 4051 Northwest Pacific cyclone tracks from 1884 to 2021. Each curve represents a full cyclone process, and the color changes gradually from red to yellow with time. (b) and (d) Cyclone trajectories obtained using revised version of the trajectory boxplot (\citeauthor{yao2020trajectory} \citeyear{yao2020trajectory}). (c) and (e) Wind speed polar plot, with wind speed as the radius and wind speed direction as the angle. The black and red represent the median and outliers, respectively, and purple, magenta, and pink indicate the first, second, and third quartile curves, respectively.}
   \label{wp_cluster}
\end{figure}

A cyclone (\citeauthor{American2000} \citeyear{American2000}) is a large-scale air mass that rotates around an intense center of low-atmospheric-pressure center. The word ``tropical'' in the term ``tropical cyclone'' refers to the geographical origin of these systems, which form almost exclusively over tropical seas. 
The rising and cooling of the swirling air results in formation of clouds and precipitation. Although most cyclones have a life cycle of 3-7 days (\citeauthor{australia} \citeyear{australia}), some systems can last for weeks if the environment is favorable. The global tropical cyclones tracks from version 4 of the International Best Track Archive for Climate Stewardship (IBTrACS v04, \citeauthor{knapp2010international} \citeyear{knapp2010international}) are available \href{https://data.humdata.org/dataset/archive-of-global-tropical-cyclone-tracks-1980-may-2019}{online}. 4051 Northwest Pacific (see Figure \ref{wp_cluster} (a)) cyclone tracks between 1884 and 2021 are extracted for clustering analysis. The data are usually measured every three hours. The time of generation and disappearance are aligned to be 0 and 1, respectively, for each trajectory.

 Using RTLP, $\theta$ is obtained to be $0.22$ according to the maximum silhouette value when $\theta$ changes from 0.01 to 0.25 in steps of 0.01 and with $p_m = 0.1$ and $\alpha = 0.85$. Two clusters were obtained according to the geographic trajectories of the cyclone centers; the first cluster has 3417 curves, and the second one has 627 curves. The clustering pattern is consistent with the pattern reported by \citeauthor{misumi2019multivariate} (\citeyear{misumi2019multivariate}). Cyclones in the first cluster (Figures \ref{wp_cluster} (b) and (c)) mainly originate from the ocean to the east of the Philippines islands and sweep into inland Asia. Usually, they end in either eastern or southwestern China, blocked by a high-pressure system from the Chinese mainland. Cyclones in the second cluster (Figures~\ref{wp_cluster} (d) and (e)) barely pass by mainland Asia, move toward North America, and disappear somewhere in the North Pacific. Seven curves that cannot be recognized in any cluster remain as outliers. They mainly show abnormal patterns or originate from a location far from the Philippines. 

 Figure \ref{wp_cluster} also maps the cyclone tracks and their storm speed information per cluster. In the storm speed polar plot, the storm direction is taken as the angle and the storm speed norm as the radius. The storm moves in the direction pointed by the vector, in degrees east of north from 0 to 360 degrees. In addition, different colors are used to denote the number of neighbours inside a cluster, with black for the median, and purple, magenta, and pink to indicate the first, second, and third quartile regions, respectively, and red to indicate outliers. 

The medians show the typical tracks in a cluster. The median in the first cluster (see Figure \ref{wp_cluster} (b)) is directed towards the northwest (direction between 270 and 360 degrees) with a storm speed of almost 10 knots, which is common for most cyclones in the first cluster. Usually, the wind speeds (i.e., the radius of the polar plot) of cyclones traveling inland decreases slowly because of reduction in the power of the cyclones. However, the median in the second cluster (see Figure \ref{wp_cluster} (d)) first shifts northwest and then turns eastwards (direction between 0 and 180 degrees), with the storm speed (i.e., the radius of the polar plot) increasing after the cyclones steer eastward. The cyclones in the second cluster that do not move inland but move to the North Pacific are influenced by the northward shift of high pressure over the Pacific (\citeauthor{japan} \citeyear{japan}). 
\begin{figure}[ht!]
    \centering
    \includegraphics[height=5.9cm,width=0.9\textwidth]{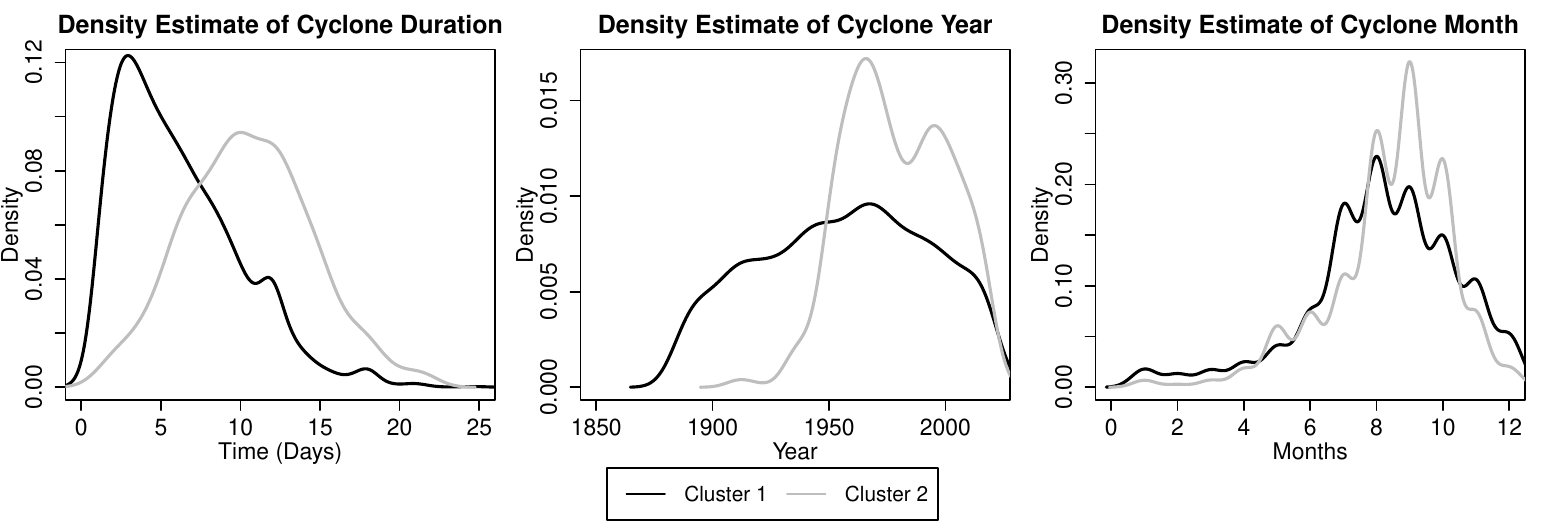}
    \caption{Extra time information from clusters 1 and 2 in the Northwest Pacific cyclones. The panels display the nonparametric density estimates of the lasting days, occurrence year, and occurrence month. The first cluster is labeled in black, and the second is in grey.}
    \label{extra_time_wp}
\end{figure}

Besides location and storm speed data, the lasting days and time of occurrence (Figure~\ref{extra_time_wp}) of the cyclones are used to analyze the difference between the clusters. The first group of cyclones mostly lasts less than ten days, and the second group mainly lasts between five and fifteen days. On average, cyclones from the first cluster (Figure \ref{wp_cluster} (b)) travel a shorter distance than those from the second cluster (Figure \ref{wp_cluster} (d)). The frequency of cyclones in both groups increased starting from the last half of the 1940s to around 1960 (\citeauthor{aoki1985climatological} \citeyear{aoki1985climatological}), especially that of the cyclones that passed by Japan and Korea and reached the Sea of Okhotsk. A remarkable decreasing trend in the cyclone frequency is noted between the late 1960s to around 1980. Subsequently, the cyclone frequency remained constant since the 1980s. Figure \ref{extra_time_wp} shows that cyclones belonging to the second cluster mainly occurred after the 1950s; however, the ratio of the frequency of cyclones traveling to mainland Asia and directed to the North Pacific is around two versus one after 1950. High pressure (\citeauthor{japan} \citeyear{japan}) in the Sea of Okhotsk in eastern Russia and low pressure to the east of the Philippines have induced shifts in the direction of cyclones since the 1950s. Cyclones in the first cluster mainly occur from July to November, whereas those in the second cluster are concentrated between August and October. The increasing frequency of the cyclones is assumed to be related to global warming (\citeauthor{van2006impacts} \citeyear{van2006impacts}), which has also influenced the rise of seawater levels and surging tides smacked in low-lying areas.

\vspace{-.3cm}

\section{Discussion}\label{sec5}
\vspace{-.2cm}
An ETD measure was defined for complete and sparse multivariate functional data. With ETD, two algorithms (two-layer partition clustering, and cluster and outlier recognition) were proposed and summarized as robust two-layer partition (RTLP) clustering. In addition, classical clustering methods such as DBSCAN, agglomerative hierarchical clustering, and $K$-medoids were naturally extended to sparse multivariate functional cases based on ETD.

The idea of RTLP is as follows. Neighbours of a curve in the set and the core in the set given a $\theta$-quantile  are defined for the algorithm application. First, the two-layer partition clustering algorithm merges curves into disjoint groups based on the concepts of neighbours and the core, and merges these groups into a set of disjoint clusters. Next, the cluster and outlier recognition algorithm determines the final primary clusters and outliers. The time complexity of the whole algorithm majorly lies in the calculation of the distance across $N$ curves, which requires $\mathcal{O}(N^2)$ time; the time complexity can be reduced through parallel computation. Regarding parameter selection, the optimal $\theta$ is selected as the parameter with the highest revised average silhouette value. 

 Using simulations, the clustering results of RTLP with three other ETD-based methods were compared: DBSCAN, agglomerative hierarchical methods, $K$-medoids, and the model-based funHDDC. The RTLP method exhibited excellent performance in terms of both clustering curves and outlier detection; the performance decreased in the following order: the RTLP method, DBSCAN, agglomerative hierarchical clustering and $K$-medoid clustering, and finally, the funHDDC algorithm. The running time of ETD-based methods was stable for a fixed sample size, and it decreased as the time grid became sparser. Overall, RTLP has a significant advantage in terms of clustering precision regardless of the outliers and the time sparseness settings.

In future works, the RTLP and current clustering methods can be applied to more general sparse (multivariate) functional data. First, the location of missing values can be not completely random and the observed time grid may carry information relevant to the phenomenon under observation (more types of sparseness are described in \citeauthor{qu2022sparse} \citeyear{qu2022sparse}); moreover, the measurement can be observed for some variables but could be missing for other variables for one index, which requires a more general distance measure. It should be noted that although the current framework transforms the curves under observation to the distance, the relevance between the missing locations and missing observations may be worth further analysis. In addition, the aforementioned clustering methods can be applied to multivariate nonfunctional data. The bridge between the RTLP clustering method and multivariate nonfunctional data can be the revised $L^p$ metric such as $p=1$, $p=2$, and $p=\infty$. Another research direction is to cluster images or surfaces (\citeauthor{chen2005clue} \citeyear{chen2005clue}, \citeauthor{genton2014surface} \citeyear{genton2014surface}); this would require defining the distance between the images.

\section*{Acknowledgments}
We thank the Editor, Associate Editor, and two anonymous reviewers for constructive comments that helped improve the paper. This publication is based upon work supported by the King Abdullah University of Science and Technology (KAUST).

\appendix
\section*{Appendix}
\textbf{Proof of Theorem 1}: 

1. $d(\bm{Y}_m,\bm{Y}_n)\geq 0$ from Eqn (\ref{d}).

2. If $\bm{Y}_m=\bm{Y}_n$, then $d(\bm{Y}_m,\bm{Y}_n)= 0$ from Eqn (\ref{d}).
~\\If $d(\bm{Y}_m,\bm{Y}_n)= 0$, i.e., $\sup\limits_{t \in \mathcal{T}}\bigg[\sqrt{\sum_{l=1}^p \Big\{Y_m^{(l)}(t)-Y_n^{(l)}(t)\Big\}^2}\bigg]=0$, then $\sum_{l=1}^p \Big\{Y_m^{(l)}(t)-Y_n^{(l)}(t)\Big\}^2=0$ $\forall t \in \mathcal{T}$. Hence $\bm{Y}_m(t)=\bm{Y}_n(t)$ for $t \in \mathcal{T}$.

3. $d(\bm{Y}_m,\bm{Y}_n)= d(\bm{Y}_n,\bm{Y}_m)$ from Eqn (\ref{d}).

4. Write $\sum_{l=1}^p\Big\{Y_m^{(l)}(t)-Y_n^{(l)}(t)\Big\}^2 = \sum_{l=1}^p\Big\{Y_m^{(l)}(t)-Y_o^{(l)}(t)+Y_o^{(l)}(t)-Y_n^{(l)}(t)\Big\}^2 = \sum_{l=1}^p\Big\{Y_m^{(l)}(t)-Y_o^{(l)}(t)\Big\}^2+\sum_{l=1}^p\Big\{Y_o^{(l)}(t)-Y_n^{(l)}(t)\Big\}^2 + \sum_{l=1}^p2\Big\{Y_m^{(l)}(t)-Y_o^{(l)}(t)\Big\}\Big\{Y_o^{(l)}(t)-Y_n^{(l)}(t)\Big\}$ $\forall~t \in \mathcal{T}$. In addition,  $\sum_{l=1}^p\Big\{Y_m^{(l)}(t)-Y_o^{(l)}(t)\Big\}\Big\{Y_o^{(l)}(t)-Y_n^{(l)}(t)\Big\} \leq \sqrt{\sum_{l=1}^p \Big\{Y_m^{(l)}(t)-Y_o^{(l)}(t)\Big\}^2}\cdot \sqrt{\sum_{l=1}^p(Y_o^{(l)}(t)-Y_n^{(l)}(t))^2}$. Hence, it is obtained that $\sum_{l=1}^p\Big\{Y_m^{(l)}(t)-Y_n^{(l)}(t)\Big\}^2 \leq \sum_{l=1}^p\Big\{Y_m^{(l)}(t)-Y_o^{(l)}(t)\Big\}^2+\sum_{l=1}^p\Big\{Y_o^{(l)}(t)-Y_n^{(l)}(t)\Big\}^2 + 2\sqrt{\sum_{l=1}^p \Big\{Y_n^{(l)}(t)-Y_o^{(l)}(t)\Big\}^2}\cdot \sqrt{\sum_{l=1}^p\Big\{Y_o^{(l)}(t)-Y_n^{(l)}(t)\Big\}^2}$. This equals to $\sum_{l=1}^p\Big\{Y_m^{(l)}(t)-Y_n^{(l)}(t)\Big\}^2 \leq \Big[\sqrt{\sum_{l=1}^p\Big\{Y_m^{(l)}(t)-Y_o^{(l)}(t)\Big\}^2}+\sqrt{\sum_{l=1}^p\Big\{Y_o^{(l)}(t)-Y_n^{(l)}(t)\Big\}^2}\Big]^2$. Hence, $\sqrt{\sum_{l=1}^p\Big\{Y_m^{(l)}(t)-Y_n^{(l)}(t)\Big\}^2} \leq \sqrt{\sum_{l=1}^p\Big\{Y_m^{(l)}(t)-Y_o^{(l)}(t)\Big\}^2}+\sqrt{\sum_{l=1}^p\Big\{Y_o^{(l)}(t)-Y_n^{(l)}(t)\Big\}^2}$. The inequality holds after taking the supremum norm. By combining the inequality relation with Eqn (\ref{d}), it is obtained that $d(\bm{Y}_m, \bm{Y}_n)\leq d(\bm{Y}_m, \bm{Y}_o) + d(\bm{Y}_o, \bm{Y}_n)$. $\hfill\Box$

\textbf{Proof of Theorem 2}: 

From Assumption \ref{time}, $g(t)$ is differentiable in $\mathcal{T}$, then for each $l=2, \ldots, T_n-1$, there exists a $t\in [t_{n, l}, t_{n, l+1}]$ such that $G(t_{n,l+1})-G(t_{n,l})=g(t)(t_{n,l+1}-t_{n,l})$, which is equivalent to $\frac{l+1}{T_n}-\frac{l}{T_n}=g(t)(t_{n,l+1}-t_{n,l})$. Since $\inf\limits_{t \in \mathcal{T}} g(t) > 0$, when $T_n \to \infty$, $t_{n,l+1}-t_{n,l}=\frac{1}{g(t)\cdot T_n} \to 0$. Sampling points grow dense when $T_n \to \infty$ for $n=1,\ldots, N$. That is, $\max\limits_{l=1, \ldots, T_n - 1}|t_{n,l+1}-t_{n,l}| \to 0$. Recall that $d(\bm{Y}_m,\bm{Y}_n): = \sup\limits_{t \in \mathcal{T}} \biggl[\sqrt{\sum_{l=1}^p \Big\{Y_m^{(l)}(t)-Y_n^{(l)}(t)\Big\}^2}\biggl]$, and $d(\widetilde{\bm{Y}}_m, \widetilde{\bm{Y}}_n):=\max\limits_{k = 1, \ldots, T}\bigg[\sqrt{\sum_{l=1}^p\Big\{\widetilde{Y}^{(l)}_{m}(st_k)-\widetilde{Y}^{(l)}_{n}(st_k)\Big\}^2}\bigg]$.

Then, the proof is split into two parts. 

First, prove ${ d(\widetilde{\bm{Y}}_m, \widetilde{\bm{Y}}_n) \xrightarrow{a.s.}{} \max\limits_{k = 1, \ldots, T}\Biggl[\sqrt{\sum_{l=1}^p\Big\{{Y}^{(l)}_{m}(st_k)-{Y}^{(l)}_{n}(st_k)\Big\}^2}\Biggr]}$; see part A.

Second, prove ${ \Biggl[\sqrt{\sum_{l=1}^p\Big\{{Y}^{(l)}_{m}(st_k)-{Y}^{(l)}_{n}(st_k)\Big\}^2}\Biggr] \xrightarrow{a.s.}{} \max\limits_{t \in [st_{k}, st_{k+1}]}\Biggl[\sqrt{\sum_{l=1}^p\Big\{{Y}^{(l)}_{m}(t)-{Y}^{(l)}_{n}(t)\Big\}^2}\Biggr] }$ for $k=1, \ldots, T$; see part B.

Then, it is easy to show { $d(\widetilde{\bm{Y}}_m, \widetilde{\bm{Y}}_n) \xrightarrow{a.s.}{} d(\bm{Y}_m,\bm{Y}_n)$} under Assumption \ref{time} and $T_n\to \infty$ for $n=1,\ldots, N$.

$d(\bm{Y}_m,\bm{Y}_n)=\max\limits_{k=1, \ldots, T -1}\Biggl(\max\limits_{t \in [st_{k}, st_{k+1}]}\Biggl[\sqrt{\sum_{l=1}^p\Big\{{Y}^{(l)}_{m}(t)-{Y}^{(l)}_{n}(t)\Big\}^2}\Biggr]\Biggr)$ because $\mathcal{T}$ is a compact set. 

A. Assumption \ref{time} implies $|t_{n, j+1}-t_{n, j}| \to 0$ as $T_n \to \infty$ $\forall~ j = 1, \ldots, T_n - 1$. For $n = 1, \ldots, N$, and for $k = 1, \ldots T$, there exists $j=1, \ldots, T_n - 1$ so that $st_k \in[t_{n,j}, t_{n,j+1}]$.~According to Definition \ref{interpolation}, $\widetilde{t}_{n,k}=\argmin\limits_{l=j, j+1}f(t_{n,l}):=\{t_{n,l}:|t_{n,l}-st_k|\}$. Then, $|\widetilde{t}_{n,k} - st_k| \leq |t_{n, j+1}-t_{n, j}|$. According to the continuity of $\bm{Y}_n$, for $l = 1, \ldots, p$, $|\widetilde{t}_{n,k} - st_k| \to 0$ leads to $|\widetilde{{Y}}^{(l)}_n(st_k)-{Y}^{(l)}_n(st_k)| = |{Y}^{(l)}_n(\widetilde{t}_{n,k}) - {Y}^{(l)}_n(st_k)| \to 0$ as $T_n \to \infty$. Similarly, for $l = 1, \ldots, p$, $|\widetilde{t}_{m,k} - st_k| \to 0$ leads to $|\widetilde{{Y}}^{(l)}_m(st_k)-{Y}^{(l)}_m(st_k)| = |{Y}^{(l)}_m(\widetilde{t}_{m,k}) - {Y}^{(l)}_m(st_k)| \to 0$ as $T_m \to \infty$. Then, $\sum_{l=1}^p\{\widetilde{Y}_m^{(l)}(st_k)-\widetilde{Y}_n^{(l)}(st_k)\}^2=\sum_{l=1}^p\{\widetilde{Y}_m^{(l)}(st_k)-Y_m^{(l)}(st_k)+Y_m^{(l)}(st_k)-\widetilde{Y}_n^{(l)}(st_k)+Y_n^{(l)}(st_k)-Y_n^{(l)}(st_k)\}^2=\sum_{l=1}^p\Biggl[\{\widetilde{Y}_m^{(l)}(st_k)-Y_m^{(l)}(st_k)\}^2 + \{\widetilde{Y}_n^{(l)}(st_k)-Y_n^{(l)}(st_k)\}^2+\{{Y}_m^{(l)}(st_k)-{Y}_n^{(l)}(st_k)\}^2+2\{\widetilde{Y}_m^{(l)}(st_k)-Y_m^{(l)}(st_k)\}\{\widetilde{Y}_n^{(l)}(st_k)-Y_n^{(l)}(st_k)\}+2\{\widetilde{Y}_m^{(l)}(st_k)-Y_m^{(l)}(st_k)\}\{{Y}_m^{(l)}(st_k)-{Y}_n^{(l)}(st_k)\}+2\{\widetilde{Y}_n^{(l)}(st_k)-Y_n^{(l)}(st_k)\}\{{Y}_m^{(l)}(st_k)-{Y}_n^{(l)}(st_k)\}\Biggr]\xrightarrow{a.s.}{} \sum_{l=1}^p\{{Y}_m^{(l)}(st_k)-{Y}_n^{(l)}(st_k)\}^2$. 

B. $\sum_{l=1}^p\{{Y}_m^{(l)}(st_k)-{Y}_n^{(l)}(st_k)\}^2=\sum_{l=1}^p\{{Y}_m^{(l)}(st_k)-{Y}_m^{(l)}(t)+Y_m^{(l)}(t)-Y_n^{(l)}(t)+Y_n^{(l)}(t)-{Y}_n^{(l)}(st_k)\}^2=\sum_{l=1}^p[\{{Y}_m^{(l)}(st_k)-{Y}_m^{(l)}(t)\}+\{Y_n^{(l)}(t)-{Y}_n^{(l)}(st_k)\}+\{Y_m^{(l)}(t)-Y_n^{(l)}(t)\}]^2$ for $t \in [st_k, st_k+1]$. Since $\mathcal{T}$ is a compact set, $|t-st_k|\leq |st_{k+1}-st_k|:=|\mathcal{T}|/T_n \to 0$ as $T_n \to \infty$ for $n=1,\ldots, N$. Then, under the continuity of $\bm{Y}_m$ and $\bm{Y}_n$, $|{Y}_m^{(l)}(st_k)-{Y}_m^{(l)}(t)| \to 0$ and $|Y_n^{(l)}(t)-{Y}_n^{(l)}(st_k)| \to 0$. Next, $\sqrt{\sum_{l=1}^p\Big\{{Y}_m^{(l)}(st_k)-{Y}_n^{(l)}(st_k)\Big\}^2} \to \sqrt{\sum_{l=1}^p\Big\{Y_m^{(l)}(t)-Y_n^{(l)}(t)\Big\}^2}$ for $t \in [st_k, st_k+1]$. Hence, $\sqrt{\sum_{l=1}^p\Big\{{Y}_m^{(l)}(st_k)-{Y}_n^{(l)}(st_k)\Big\}^2} \to \max\limits_{t \in [st_k, st_{k+1}]}\Biggl[\sqrt{\sum_{l=1}^p\Big\{Y_m^{(l)}(t)-Y_n^{(l)}(t)\Big\}^2}\Biggr]$, which concludes the proof.   \hfill $\Box$

\baselineskip=15pt
\bibliographystyle{apalike2}
\bibliography{ref}

\begin{thebibliography}{}

\bibitem[Abraham et~al., 2003]{abraham2003unsupervised}
Abraham, C., Cornillon, P.-A., Matzner-L{\o}ber, E., \& Molinari, N. (2003).
\newblock Unsupervised curve clustering using {B}-splines.
\newblock {\em Scandinavian Journal of Statistics}, 30(3), 581--595.

\bibitem[Akaike, 1974]{akaike1974new}
Akaike, H. (1974).
\newblock A new look at the statistical model identification.
\newblock {\em IEEE Transactions on Automatic Control}, 19(6), 716--723.

\bibitem[Al~Abid, 2014]{al2014novel}
Al~Abid, F.~B. (2014).
\newblock A novel approach for {PAM} clustering method.
\newblock {\em International Journal of Computer Applications}, 86(17), 1--5.

\bibitem[Albert-Smet et~al., 2022]{albert2021band}
Albert-Smet, J., Torrente, A., \& Romo, J. (2022).
\newblock Band depth based initialization of $k$-means for functional data
  clustering.
\newblock {\em Advances in Data Analysis and Classification}, to appear.

\bibitem[{American Meteorological Society}, 2000]{American2000}
{American Meteorological Society} (2000).
\newblock Glossary of meteorology: Cyclonic circulation.
\newblock https://glossary.ametsoc.org/wiki/Cyclonic{\_}circulation.
\newblock accessed: 27.08.2021.

\bibitem[Antoniadis et~al., 2013]{antoniadis2013clustering}
Antoniadis, A., Brossat, X., Cugliari, J., \& Poggi, J.-M. (2013).
\newblock Clustering functional data using wavelets.
\newblock {\em International Journal of Wavelets, Multiresolution and
  Information Processing}, 11(01), 1350003.

\bibitem[AOKI, 1985]{aoki1985climatological}
AOKI, T. (1985).
\newblock A climatological study of typhoon formation and typhoon visit to
  {J}apan.
\newblock {\em Meteorology and Geophysics}, 36, 61--118.

\bibitem[Balcan et~al., 2014]{balcan2014robust}
Balcan, M.-F., Liang, Y., \& Gupta, P. (2014).
\newblock Robust hierarchical clustering.
\newblock {\em The Journal of Machine Learning Research}, 15(1), 3831--3871.

\bibitem[Batool \& Hennig, 2021]{batool2021clustering}
Batool, F. \& Hennig, C. (2021).
\newblock Clustering with the average silhouette width.
\newblock {\em Computational Statistics \& Data Analysis}, 158, 107190.

\bibitem[Biernacki et~al., 2000]{biernacki2000assessing}
Biernacki, C., Celeux, G., \& Govaert, G. (2000).
\newblock Assessing a mixture model for clustering with the integrated
  completed likelihood.
\newblock {\em IEEE Transactions on Pattern Analysis and Machine Intelligence},
  22(7), 719--725.

\bibitem[Boull{\'e}, 2012]{boulle2012functional}
Boull{\'e}, M. (2012).
\newblock Functional data clustering via piecewise constant nonparametric
  density estimation.
\newblock {\em Pattern Recognition}, 45(12), 4389--4401.

\bibitem[Bouveyron \& Jacques, 2011]{bouveyron2011model}
Bouveyron, C. \& Jacques, J. (2011).
\newblock Model-based clustering of time series in group-specific functional
  subspaces.
\newblock {\em Advances in Data Analysis and Classification}, 5(4), 281--300.

\bibitem[Centofanti et~al., 2021]{centofanti2021sparse}
Centofanti, F., Lepore, A., \& Palumbo, B. (2021).
\newblock Sparse and smooth functional data clustering.
\newblock {\em arXiv preprint arXiv:2103.15224}.

\bibitem[Chen et~al., 2017]{chen2017delineating}
Chen, Y., Liu, X., Li, X., Liu, X., Yao, Y., Hu, G., Xu, X., \& Pei, F. (2017).
\newblock Delineating urban functional areas with building-level social media
  data: A dynamic time warping ({DTW}) distance based {K}-medoids method.
\newblock {\em Landscape and Urban Planning}, 160, 48--60.

\bibitem[Chen et~al., 2005]{chen2005clue}
Chen, Y., Wang, J.~Z., \& Krovetz, R. (2005).
\newblock Clue: Cluster-based retrieval of images by unsupervised learning.
\newblock {\em IEEE Transactions on Image Processing}, 14(8), 1187--1201.

\bibitem[Chiou \& Li, 2007]{chiou2007functional}
Chiou, J.-M. \& Li, P.-L. (2007).
\newblock Functional clustering and identifying substructures of longitudinal
  data.
\newblock {\em Journal of the Royal Statistical Society: Series B (Statistical
  Methodology)}, 69(4), 679--699.

\bibitem[Cuesta-Albertos \& Fraiman, 2007]{cuesta2007impartial}
Cuesta-Albertos, J.~A. \& Fraiman, R. (2007).
\newblock Impartial trimmed {K}-means for functional data.
\newblock {\em Computational Statistics \& Data Analysis}, 51(10), 4864--4877.

\bibitem[Dai \& Genton, 2018]{dai2018multivariate}
Dai, W. \& Genton, M.~G. (2018).
\newblock Multivariate functional data visualization and outlier detection.
\newblock {\em Journal of Computational and Graphical Statistics}, 27(4),
  923--934.

\bibitem[Dai \& Genton, 2019]{dai2019directional}
Dai, W. \& Genton, M.~G. (2019).
\newblock Directional outlyingness for multivariate functional data.
\newblock {\em Computational Statistics \& Data Analysis}, 131, 50--65.

\bibitem[Day \& Edelsbrunner, 1984]{day1984efficient}
Day, W.~H. \& Edelsbrunner, H. (1984).
\newblock Efficient algorithms for agglomerative hierarchical clustering
  methods.
\newblock {\em Journal of Classification}, 1(1), 7--24.

\bibitem[Ester et~al., 1996]{ester1996density}
Ester, M., Kriegel, H.-P., Sander, J., \& Xu, X. (1996).
\newblock Density-based spatial clustering of applications with noise.
\newblock In {\em Int. Conf. Knowledge Discovery and Data Mining}, volume 240
  (pp.\~6).

\bibitem[Ferreira \& Hitchcock, 2009]{ferreira2009comparison}
Ferreira, L. \& Hitchcock, D.~B. (2009).
\newblock A comparison of hierarchical methods for clustering functional data.
\newblock {\em Communications in Statistics-Simulation and Computation}, 38(9),
  1925--1949.

\bibitem[Floriello \& Vitelli, 2017]{floriello2017sparse}
Floriello, D. \& Vitelli, V. (2017).
\newblock Sparse clustering of functional data.
\newblock {\em Journal of Multivariate Analysis}, 154, 1--18.

\bibitem[Gagolewski et~al., 2016]{gagolewski2016genie}
Gagolewski, M., Bartoszuk, M., \& Cena, A. (2016).
\newblock Genie: A new, fast, and outlier-resistant hierarchical clustering
  algorithm.
\newblock {\em Information Sciences}, 363, 8--23.

\bibitem[Garc{\'\i}a et~al., 2015]{garcia2015k}
Garc{\'\i}a, M. L.~L., Garc{\'\i}a-R{\'o}denas, R., \& G{\'o}mez, A.~G. (2015).
\newblock K-means algorithms for functional data.
\newblock {\em Neurocomputing}, 151, 231--245.

\bibitem[Genton et~al., 2014]{genton2014surface}
Genton, M.~G., Johnson, C., Potter, K., Stenchikov, G., \& Sun, Y. (2014).
\newblock Surface boxplots.
\newblock {\em Stat}, 3(1), 1--11.

\bibitem[Giacofci et~al., 2013]{giacofci2013wavelet}
Giacofci, M., Lambert-Lacroix, S., Marot, G., \& Picard, F. (2013).
\newblock Wavelet-based clustering for mixed-effects functional models in high
  dimension.
\newblock {\em Biometrics}, 69(1), 31--40.

\bibitem[Gneiting et~al., 2010]{gneiting2010matern}
Gneiting, T., Kleiber, W., \& Schlather, M. (2010).
\newblock Mat{\'e}rn cross-covariance functions for multivariate random fields.
\newblock {\em Journal of the American Statistical Association}, 105(491),
  1167--1177.

\bibitem[Hartigan \& Wong, 1979]{hartigan1979algorithm}
Hartigan, J.~A. \& Wong, M.~A. (1979).
\newblock Algorithm {AS} 136: A {K}-means clustering algorithm.
\newblock {\em Journal of the Royal Statistical Society. Series C (Applied
  Statistics)}, 28(1), 100--108.

\bibitem[Horv{\'a}th \& Kokoszka, 2012]{horvath2012inference}
Horv{\'a}th, L. \& Kokoszka, P. (2012).
\newblock {\em Inference for Functional Data with Applications}, volume 200.
\newblock Springer Science \& Business Media.

\bibitem[Hsing \& Eubank, 2015]{hsing2015theoretical}
Hsing, T. \& Eubank, R. (2015).
\newblock {\em Theoretical Foundations of Functional Data Analysis, with an
  Introduction to Linear Operators}, volume 997.
\newblock John Wiley \& Sons.

\bibitem[Hubert \& Arabie, 1985]{hubert1985comparing}
Hubert, L. \& Arabie, P. (1985).
\newblock Comparing partitions.
\newblock {\em Journal of Classification}, 2(1), 193--218.

\bibitem[Hubert et~al., 2015]{hubert2015multivariate}
Hubert, M., Rousseeuw, P.~J., \& Segaert, P. (2015).
\newblock Multivariate functional outlier detection.
\newblock {\em Statistical Methods \& Applications}, 24(2), 177--202.

\bibitem[Ieva et~al., 2011]{ieva2011multivariate}
Ieva, F., Paganoni, A.~M., Pigoli, D., \& Vitelli, V. (2011).
\newblock Multivariate functional clustering for the analysis of {E}{C}{G}
  curves morphology.
\newblock In {\em Cladag 2011 (8th International Meeting of the Classification
  and Data Analysis Group)}  (pp.\ 1--4).

\bibitem[Ieva et~al., 2013]{ieva2013multivariate}
Ieva, F., Paganoni, A.~M., Pigoli, D., \& Vitelli, V. (2013).
\newblock Multivariate functional clustering for the morphological analysis of
  electrocardiograph curves.
\newblock {\em Journal of the Royal Statistical Society: Series C (Applied
  Statistics)}, 62(3), 401--418.

\bibitem[Jacques \& Preda, 2013]{jacques2013funclust}
Jacques, J. \& Preda, C. (2013).
\newblock Funclust: A curves clustering method using functional random
  variables density approximation.
\newblock {\em Neurocomputing}, 112, 164--171.

\bibitem[Jacques \& Preda, 2014a]{jacques2014functional}
Jacques, J. \& Preda, C. (2014a).
\newblock Functional data clustering: A survey.
\newblock {\em Advances in Data Analysis and Classification}, 8(3), 231--255.

\bibitem[Jacques \& Preda, 2014b]{jacques2014model}
Jacques, J. \& Preda, C. (2014b).
\newblock Model-based clustering for multivariate functional data.
\newblock {\em Computational Statistics \& Data Analysis}, 71, 92--106.

\bibitem[James \& Sugar, 2003]{james2003clustering}
James, G.~M. \& Sugar, C.~A. (2003).
\newblock Clustering for sparsely sampled functional data.
\newblock {\em Journal of the American Statistical Association}, 98(462),
  397--408.

\bibitem[Jeong et~al., 2016]{jeong2016data}
Jeong, M.-H., Cai, Y., Sullivan, C.~J., \& Wang, S. (2016).
\newblock Data depth based clustering analysis.
\newblock In {\em Proceedings of the 24th ACM SIGSPATIAL International
  Conference on Advances in Geographic Information Systems}  (pp.\ 1--10).

\bibitem[Kaufman \& Rousseeuw, 1990]{kaufman1990partitioning}
Kaufman, L. \& Rousseeuw, P.~J. (1990).
\newblock Partitioning around medoids (program pam).
\newblock {\em Finding Groups in Data: An Introduction to Cluster Analysis},
  344, 68--125.

\bibitem[Kelley, 1955]{Kelley1955general}
Kelley, J.~L. (1955).
\newblock {\em General Topology}.
\newblock D. Van Nostrand.

\bibitem[Knapp et~al., 2010]{knapp2010international}
Knapp, K.~R., Kruk, M.~C., Levinson, D.~H., Diamond, H.~J., \& Neumann, C.~J.
  (2010).
\newblock The international best track archive for climate stewardship
  (ibtracs) unifying tropical cyclone data.
\newblock {\em Bulletin of the American Meteorological Society}, 91(3),
  363--376.

\bibitem[Kneip et~al., 2000]{kneip2000curve}
Kneip, A., Li, X., MacGibbon, K., \& Ramsay, J. (2000).
\newblock Curve registration by local regression.
\newblock {\em Canadian Journal of Statistics}, 28(1), 19--29.

\bibitem[Li et~al., 2007]{li2007noise}
Li, Z., Liu, J., Chen, S., \& Tang, X. (2007).
\newblock Noise robust spectral clustering.
\newblock {\em 2007 IEEE 11th International Conference on Computer Vision},
  (pp.\ 1--8).

\bibitem[Liu et~al., 2012]{liu2012correlation}
Liu, X., Zhu, X.-H., Qiu, P., \& Chen, W. (2012).
\newblock A correlation-matrix-based hierarchical clustering method for
  functional connectivity analysis.
\newblock {\em Journal of Neuroscience Methods}, 211(1), 94--102.

\bibitem[Marron et~al., 2014]{10.1214/14-EJS901}
Marron, J.~S., Ramsay, J.~O., Sangalli, L.~M., \& Srivastava, A. (2014).
\newblock {Statistics of time warpings and phase variations}.
\newblock {\em Electronic Journal of Statistics}, 8(2), 1697 -- 1702.

\bibitem[Meng et~al., 2018]{meng2018new}
Meng, Y., Liang, J., Cao, F., \& He, Y. (2018).
\newblock A new distance with derivative information for functional {K}-means
  clustering algorithm.
\newblock {\em Information Sciences}, 463, 166--185.

\bibitem[Misumi et~al., 2019]{misumi2019multivariate}
Misumi, T., Matsui, H., \& Konishi, S. (2019).
\newblock Multivariate functional clustering and its application to typhoon
  data.
\newblock {\em Behaviormetrika}, 46(1), 163--175.

\bibitem[Park \& Ahn, 2017]{park2017clustering}
Park, J. \& Ahn, J. (2017).
\newblock Clustering multivariate functional data with phase variation.
\newblock {\em Biometrics}, 73(1), 324--333.

\bibitem[Peng \& M{\"u}ller, 2008]{peng2008distance}
Peng, J. \& M{\"u}ller, H.-G. (2008).
\newblock Distance-based clustering of sparsely observed stochastic processes,
  with applications to online auctions.
\newblock {\em The Annals of Applied Statistics}, 2(3), 1056--1077.

\bibitem[Qu \& Genton, 2022]{qu2022sparse}
Qu, Z. \& Genton, M.~G. (2022).
\newblock Sparse functional boxplots for multivariate curves.
\newblock {\em Journal of Computational and Graphical Statistics}, to appear.

\bibitem[Ramsay \& Silverman, 2005]{ramsay2005functional}
Ramsay, J.~O. \& Silverman, B.~W. (2005).
\newblock {\em Functional Data Analysis}.
\newblock Springer Series in Statistics. Springer Science \& Business Media.

\bibitem[Ronchetti, 2021]{ronchetti2021main}
Ronchetti, E. (2021).
\newblock The main contributions of robust statistics to statistical science
  and a new challenge.
\newblock {\em Special Issue on the Centenary of Metron. Metron: International
  Journal of Statistics}, 79(2), 127--135.

\bibitem[Rossini et~al., 2007]{rossini2007simple}
Rossini, A.~J., Tierney, L., \& Li, N. (2007).
\newblock Simple parallel statistical computing in {R}.
\newblock {\em Journal of {C}omputational and {G}raphical Statistics}, 16(2),
  399--420.

\bibitem[Rousseeuw, 1987]{rousseeuw1987silhouettes}
Rousseeuw, P.~J. (1987).
\newblock Silhouettes: a graphical aid to the interpretation and validation of
  cluster analysis.
\newblock {\em Journal of Computational and Applied Mathematics}, 20, 53--65.

\bibitem[Sam{\'e} et~al., 2011]{same2011model}
Sam{\'e}, A., Chamroukhi, F., Govaert, G., \& Aknin, P. (2011).
\newblock Model-based clustering and segmentation of time series with changes
  in regime.
\newblock {\em Advances in Data Analysis and Classification}, 5(4), 301--321.

\bibitem[Sangalli et~al., 2010]{sangalli2010k}
Sangalli, L.~M., Secchi, P., Vantini, S., \& Vitelli, V. (2010).
\newblock K-mean alignment for curve clustering.
\newblock {\em Computational Statistics \& Data Analysis}, 54(5), 1219--1233.

\bibitem[Schmutz et~al., 2020]{schmutz2020clustering}
Schmutz, A., Jacques, J., Bouveyron, C., Cheze, L., \& Martin, P. (2020).
\newblock Clustering multivariate functional data in group-specific functional
  subspaces.
\newblock {\em Computational Statistics}, 35, 1101--1131.

\bibitem[Schwarz, 1978]{schwarz1978estimating}
Schwarz, G. (1978).
\newblock Estimating the dimension of a model.
\newblock {\em The Annals of Statistics}, 6(2), 461--464.

\bibitem[Sheehy et~al., 2000]{sheehy2000contribution}
Sheehy, A., Gasser, T., Molinari, L., \& Largo, R.~H. (2000).
\newblock Contribution of growth phases to adult size.
\newblock {\em Annals of Human Biology}, 27(3), 281--298.

\bibitem[Shi et~al., 2021]{shi2021quantitative}
Shi, C., Wei, B., Wei, S., Wang, W., Liu, H., \& Liu, J. (2021).
\newblock A quantitative discriminant method of elbow point for the optimal
  number of clusters in clustering algorithm.
\newblock {\em EURASIP Journal on Wireless Communications and Networking},
  2021(1), 1--16.

\bibitem[{Storm Science Australia}, 2008]{australia}
{Storm Science Australia} (2008).
\newblock Tropical cyclones.
\newblock https://www.ausstormscience.com/tropical-cyclones/.
\newblock accessed: 27.08.2021.

\bibitem[Struyf et~al., 1997]{struyf1997clustering}
Struyf, A., Hubert, M., \& Rousseeuw, P. (1997).
\newblock Clustering in an object-oriented environment.
\newblock {\em Journal of Statistical Software}, 1(4), 1--30.

\bibitem[Tarpey \& Kinateder, 2003]{tarpey2003clustering}
Tarpey, T. \& Kinateder, K.~K. (2003).
\newblock Clustering functional data.
\newblock {\em Journal of Classification}, 20(1), 093--114.

\bibitem[Tibshirani et~al., 2001]{tibshirani2001estimating}
Tibshirani, R., Walther, G., \& Hastie, T. (2001).
\newblock Estimating the number of clusters in a data set via the gap
  statistic.
\newblock {\em Journal of the Royal Statistical Society: Series B}, 63(2),
  411--423.

\bibitem[Times, 2022]{japan}
Times, N.~Y. (2022).
\newblock Nature, science, animals - land, weather, volcanoes and earthquakes.
\newblock https://factsanddetails.com/japan/cat26/sub160/item856.html.
\newblock accessed: 03.01.2022.

\bibitem[Van~Aalst, 2006]{van2006impacts}
Van~Aalst, M.~K. (2006).
\newblock The impacts of climate change on the risk of natural disasters.
\newblock {\em Disasters}, 30(1), 5--18.

\bibitem[Wang \& Su, 2011]{wang2011improved}
Wang, J. \& Su, X. (2011).
\newblock An improved {K}-means clustering algorithm.
\newblock {\em 2011 IEEE 3rd International Conference on Communication Software
  and Networks}, (pp.\ 44--46).

\bibitem[Yao et~al., 2005]{yao2005functional}
Yao, F., M{\"u}ller, H.-G., \& Wang, J.-L. (2005).
\newblock Functional data analysis for sparse longitudinal data.
\newblock {\em Journal of the American Statistical Association}, 100(470),
  577--590.

\bibitem[Yao et~al., 2020]{yao2020trajectory}
Yao, Z., Dai, W., \& Genton, M.~G. (2020).
\newblock Trajectory functional boxplots.
\newblock {\em Stat}, 9(1), e289.

\end{thebibliography}







\end{document}